\documentclass[a4paper,fleqn,usenatbib]{mnras}

\usepackage{mathptmx}
\usepackage{txfonts}
\usepackage{lscape}

\usepackage[T1]{fontenc}
\usepackage{ae,aecompl}

\usepackage{graphicx}	

\newcommand\iso[2]{$^{\rm #1}$#2}

\title[NGC 6402 Abundances]{Light Element Discontinuities Suggest an Early 
Termination of Star Formation in the Globular Cluster NGC 6402 (M14)}

\author[Johnson et al.]{
Christian I. Johnson,$^{1}$\thanks{E-mail: cjohnson@cfa.harvard.edu}
Nelson Caldwell,$^{1}$
R. Michael Rich,$^{2}$
Mario Mateo,$^{3}$ and
\newauthor
John I. Bailey, III$^{4}$
\\
$^{1}$Center for Astrophysics $|$ Harvard \& Smithsonian, 60 Garden Street,  
Cambridge, MA 02138, USA\\
$^{2}$Department of Physics and Astronomy, UCLA, 430 Portola Plaza,
Box 951547, Los Angeles, CA 90095-1547, USA\\
$^{3}$Department of Astronomy, University of Michigan, Ann Arbor,
MI 48109, USA\\
$^{4}$Department of Physics, UCSB, Santa Barbara, CA 93016, USA
}

\date{Accepted XXX. Received YYY; in original form ZZZ}

\pubyear{2018}

\begin{document}
\label{firstpage}
\pagerange{\pageref{firstpage}--\pageref{lastpage}}
\maketitle

\begin{abstract}
NGC 6402 is among the most massive globular clusters in the Galaxy, but little
is known about its detailed chemical composition.  Therefore, we obtained
radial velocities and/or chemical abundances of 11 elements for 41 red giant 
branch cluster members using high resolution spectra obtained with the 
Magellan-M2FS instrument.  We find NGC 6402 to be only moderately metal-poor
with $\langle$[Fe/H]$\rangle$ = $-$1.13 dex ($\sigma$ = 0.05 dex) and to have
a mean heliocentric radial velocity of $-$61.1 km s$^{\rm -1}$ ($\sigma$ = 8.5
km s$^{\rm -1}$).  In general, NGC 6402 exhibits mean composition properties
that are similar to other inner Galaxy clusters, such as [$\alpha$/Fe] $\sim$
$+$0.3 dex, [Cr,Ni/Fe] $\sim$ 0.0 dex, and $\langle$[La/Eu]$\rangle$ = $-$0.08
dex.  Similarly, we find large star-to-star abundance variations for O, Na, Mg,
Al, and Si that are indicative of gas that experienced high temperature 
proton-capture burning.  Interestingly, we detect three distinct populations, 
but also find large gaps in the [O/Fe], [Na/Fe], and [Al/Fe] distributions that
may provide the first direct evidence of delayed formation for intermediate 
composition stars.  A qualitative enrichment model is discussed where 
clusters form stars through an early ($\la$ 5$-$10 Myr) phase, which results 
in first generation and ``extreme" composition stars, and a delayed phase 
($\ga$ 40 Myr), which results in the dilution of processed and pristine gas 
and the formation of intermediate composition stars.  For NGC 6402, the missing
intermediate composition stars suggest the delayed phase terminated 
prematurely, and as a result the cluster may uniquely preserve details of the 
chemical enrichment process.

\end{abstract}

\begin{keywords}
galaxies: star clusters: individual: NGC 6402 (M14)
\end{keywords}



\section{Introduction}

Early high resolution spectroscopic work on Galactic globular clusters 
discovered the presence of large, often correlated, star-to-star abundance 
variations for elements ranging from at least carbon through aluminum 
\citep[e.g., see reviews by][and references therein]{Kraft94,Gratton04}.  The
observed patterns, such as C-N, O-Na, and Mg-Al anti-correlations
coupled with Na-Al correlations, hinted at an origin rooted in high temperature
($>$ 40 MK) proton-capture nucleosynthesis \citep[e.g.,][]{Denisenkov90,
Langer93}.  Although originally observed in bright red giant
branch (RGB) stars in some of the nearest and most massive clusters, recent
large sample spectroscopic surveys \citep[e.g.,][]{Carretta09_gir,
Carretta09_uves} have shown that (anti-)correlated light element abundance 
variations are ubiquitous patterns found in nearly all old globular clusters 
and at all evolutionary stages \citep[e.g.,][]{Gratton01}.

Interestingly, the abundance variations tend to ``clump" into discrete groups
rather than trace a composition continuum.  The data indicate that most old
globular clusters are not monolithic populations, and instead are collections
of two or more chemically distinct groups.  The discrete nature of globular 
cluster formation and enrichment is most clearly seen in the \emph{Hubble Space
Telescope} (\emph{HST}) color-magnitude diagram (CMD) compilations of 
\citet{Piotto15} and \citet{Milone17_atlas}, but is also visible in the light 
element abundance diagrams of some spectroscopic analyses 
\citep[e.g.,][]{Carretta12_6752,Cohen12,Mucciarelli15,Johnson17_5986,
Marino17_agb,Carretta18}.  

With few exceptions\footnote{The low mass Galactic globular clusters Rup 106 
and E~3 may contain only single populations of first generation stars
\citep{Villanova13,Salinas15,Dotter18}.}, globular clusters can be broadly 
decomposed into sets of ``first generation" and ``second generation" stars.
The first generation stars are characterized as having light 
element compositions that mirror those of similar metallicity halo field stars
\citep[e.g.,][]{Sneden04} while the second generation groups tend to have
lower C, O, and Mg abundances and higher He, N, Na, and Al abundances.  Aside 
from a general consensus that second generation stars likely formed from gas 
that was processed at temperatures exceeding 40$-$80 MK 
\citep[e.g.,][]{Langer93,Prantzos07,Denissenkov15,Ventura18}, little agreement 
exists regarding the origins of the (anti-)correlated light element variations.
Various enrichment sites have been proposed, including intermediate mass 
asymptotic giant branch (AGB) stars \citep[e.g.,][]{Ventura09,Karakas14}, fast 
rotating massive stars \citep[e.g.,][]{Decressin07,Krause13}, interacting 
massive binary stars \citep{deMink09,Bastian13}, Wolf-Rayet stars 
\citep{Smith06}, white dwarf novae \citep{Maccarone12}, black hole accretion 
disks \citep{Breen18}, and supermassive stars \citep{Denissenkov14}.  However, 
these scenarios currently fail to simultaneously explain all globular cluster 
composition characteristics \citep[e.g.,][]{Bastian15,Renzini15,Bastian18}.  

Despite the theoretical difficulties in explaining the star-to-star abundance
variations, observations clearly show strong links between a cluster's 
horizontal branch (HB) extension, cluster mass, and composition range.  In 
particular, clusters possessing warmer blue HB populations and larger 
present-day masses tend to exhibit more extreme [O/Na] and [Mg/Al] abundance 
ranges \citep{Carretta07,Carretta10,Gratton10}.  A significant fraction of 
clusters with M$_{\rm V}$ $\la$ $-$8 exhibit variations in Fe-peak and 
s-process abundances as well, and some evidence indicates that many of these 
``Fe-complex" or ``Type II" clusters may be the remnant cores of former dwarf 
galaxies \citep[e.g.,][]{DaCosta16,Johnson17_6229,Marino18}.  In fact, 
\citet{Lee07} suggests that most or all clusters with significant blue HB 
populations have an accretion origin.

Following the evidence outlined above, we targeted RGB stars in the massive 
blue HB inner Galaxy cluster NGC 6402 for analysis with high resolution
spectroscopy.  Although little is known about the cluster's detailed chemical 
composition, several lines of evidence, including possible membership in a 
stellar stream \citep{Gao07}, suggest NGC 6402 may be an accreted system.  For 
instance, NGC 6402 contains a rare CH star \citep{Cote97}.  Such objects
are common in dwarf galaxies \citep[e.g.,][]{Cannon81}, but in the Milky Way
only two other clusters, $\omega$ Cen \citep{Harding62,Dickens72,Bond75} and
NGC 6426 \citep{Sharina12}, have confirmed detections.  Additionally,
\citet{ContrerasPena18} found that NGC 6402 is likely an 
Oosterhoff-intermediate cluster, which is a signature more reminiscent of 
local dwarf galaxies than Galactic globular clusters.

NGC 6402 is among the brightest globular clusters in the Milky Way,
and nearly all of the clusters with similar luminosity possess complex chemical
patterns.  For example, $\omega$ Cen \citep[e.g.,][]{Johnson10,Marino11}, 
NGC 6273 \citep{Johnson15_6273,Johnson17_6273}, and M 2 \citep{Yong14_M2,
Milone15,Lardo16} each host at least 2$-$3 groups of stars with distinct light 
and heavy element compositions, M 54 exhibits a broad metallicity distribution 
and may be the nuclear core of the Sagittarius dwarf galaxy 
\citep[e.g.,][]{Carretta10_M54}, NGC 6388 and NGC 6441 have unusually extended 
blue HBs and light element variations for their high metallicities
\citep[e.g.,][]{Carretta18_N6388,Gratton07,Tailo17}, and NGC 2419 and NGC 2808
show signatures of extremely high temperature ($>$ 100 MK) proton-capture 
burning \citep[e.g.,][]{Cohen12,Ventura12,Carretta15,Mucciarelli15,Prantzos17}.
Although the present work does not find evidence supporting a metallicity 
spread, we report that NGC 6402 maintains the aforementioned trend of massive 
clusters exhibiting peculiar composition characteristics.

\section{Observations and Data Reduction}

The spectra for this project were obtained using the Michigan-Magellan Fiber
System \citep[M2FS;][]{Mateo12} and MSpec spectrograph on the Magellan-Clay
6.5m telescope at Las Campanas Observatory.  Both spectrographs were configured
to use the 125$\mu$m slits, a four amplifier slow readout mode, 1 $\times$ 2
binning in the dispersion and spatial directions, the ``Bulge$\_$GC1" order 
blocking filters, and the echelle gratings.  This setup permits up to 48 fibers
to be placed on targets within the 29.2$\arcmin$ field-of-view and produces
spectra with a mean resolving power of R $\equiv$ $\lambda$/$\Delta$$\lambda$
$\approx$ 27,000.  The total wavelength coverage of all six orders ranged from 
$\sim$6140$-$6720 \AA.

\begin{figure}
\includegraphics[width=\columnwidth]{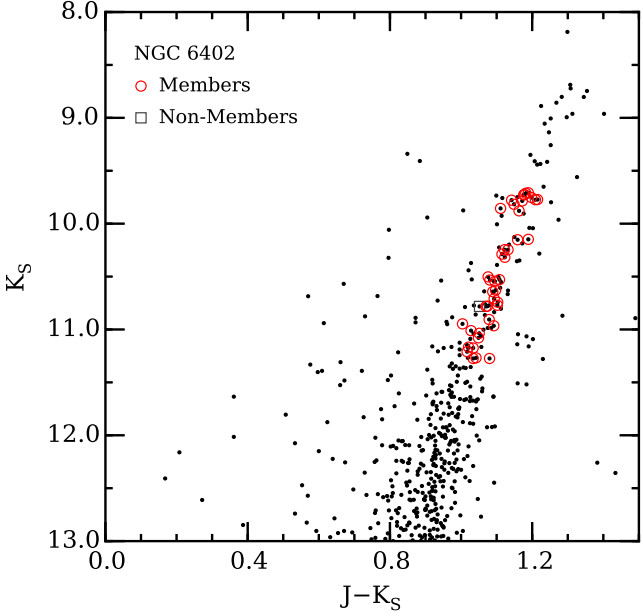}
\caption{A 2MASS color-magnitude diagram is shown for a region within
5$\arcmin$ of NGC 6402 (small black circles).  Cluster member and non-member
stars are indicated with open red circles and open grey boxes, respectively.}
\label{fig:f1}
\end{figure}

A total of 42 fibers were placed on possible RGB stars within a radius of 
$\sim$3.6$\arcmin$ from the core of NGC 6402, and an additional 6 fibers were 
placed on blank sky regions.  No targets were selected within 0.5$\arcmin$ of
the cluster core due to severe crowding restricting the placement of single
stars within the 1.2$\arcsec$ fibers.  Targets and coordinates were obtained 
using the Two Micron All Sky Survey \citep[2MASS;][]{Skrutskie06}, and 
membership probabilities were estimated using a fiducial sequence derived from 
2MASS CMDs that only included stars within 1$\arcmin$ of the cluster core.  The
final target selection is illustrated in Fig.~\ref{fig:f1} and spans 
K$_{\rm S}$ = 9.7$-$11.3 mag., which is equivalent to V = 14.7$-$15.7 mag.
The brightest RGB-tip stars were not included in order to avoid targets with 
effective temperatures (T$_{\rm eff}$) lower than $\sim$4100 K.  The target 
identifiers, J2000 coordinates, and 2MASS photometry are provided in 
Table~\ref{tab:basic_params}.

The spectra were obtained under clear but moderately humid conditions with
reasonable seeing ($\sim$1.2$\arcsec$) on 2017 September 11, 12, and 15.  The 
integration time totaled 18,000 seconds (5 hours) and was distributed over ten
1800 second exposures.  Signal-to-noise ratios (S/N) for individual exposures 
were approximately 25$-$50 per reduced pixel.

Data reduction followed the recipe outlined in \citet{Johnson15_6273}, which
primarily exploits the \emph{dohydra} routine from IRAF\footnote{IRAF is 
distributed by the National Optical Astronomy Observatory, which is operated 
by the Association of Universities for Research in Astronomy, Inc., under 
cooperative agreement with the National Science Foundation.} to carry out 
aperture identification and tracing, flat-fielding, scattered light removal, 
wavelength calibration, cosmic-ray removal, throughput correction, and 
spectrum extraction.  Basic image preparation such as overscan and bias 
correction, overscan trimming, and dark current removal was completed with
IRAF before processing with \emph{dohydra}.  The extracted sky spectra for 
each order and exposure were median combined separately and subtracted from
the object spectra.  The sky subtracted stellar spectra were continuum 
normalized, divided by a previously obtained telluric spectrum, corrected for
heliocentric velocity variations, and median combined.  The final S/N ranged 
from $\sim$75$-$130 per reduced pixel.

\section{Radial Velocities and Cluster Membership}

Radial velocities were measured for each order and exposure of every star using
the XCSAO cross-correlation routine \citep{Kurtz98}.  The velocities were
measured relative to a synthetic spectrum library spanning the temperature,
gravity, and metallicity regime probed here.  All template spectra were 
smoothed and resampled to match the observations.  Heliocentric radial velocity
corrections were calculated for each exposure using the IRAF \emph{rvcor}
routine and applied to the relative velocity measurements output from XCSAO.
The mean heliocentric corrected velocities for all stars are reported in 
Table~\ref{tab:basic_params}.  Note that the velocity errors provided in 
Table~\ref{tab:basic_params} represent the standard deviations of all 
measurements (i.e., each order and exposure) and range from 0.1$-$2.0 
km s$^{\rm -1}$.

\begin{figure}
\includegraphics[width=\columnwidth]{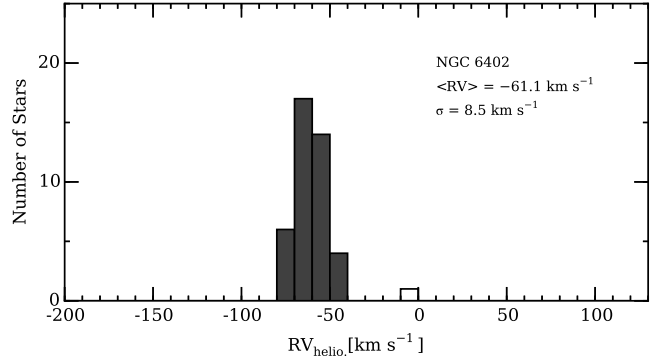}
\caption{The heliocentric radial velocity distributions of the member (filled
grey histogram) and non-member (white histogram) stars are shown using a bin
width of 10 km s$^{\rm -1}$.}
\label{fig:f2}
\end{figure}

Fig.~\ref{fig:f2} shows the heliocentric radial velocity distribution of our
targets, and indicates that the cluster members are clearly separated from
the field star population.  In fact, 41/42 stars observed for this program
are cluster members based on velocity.  The sole non-member resides at $-$7.7 
km s$^{\rm -1}$, which is consistent with the broad field distribution that 
peaks near $-$20 km s$^{\rm -1}$, as noted by \citet{Cote97}.  Ignoring the 
non-member star, we find NGC 6402 to have a mean heliocentric radial velocity
of $-$61.1 km s$^{\rm -1}$ with a dispersion of 8.5 km s$^{\rm -1}$.  Early
estimates of the cluster velocity by \citet{Webbink81} and \citet{Hesser86}
found values of $-$123 km s$^{\rm -1}$ and $-$25 km s$^{\rm -1}$, respectively,
but as pointed out by \citet{Cote97} these results likely suffered from
field star contamination and/or inaccurate measurements.  More recent velocity
estimates from \citet{Cote97}, \citet{Kimmig15}, and \citet{Baumgardt18} 
range from $-$59.5 km s$^{\rm -1}$ to $-$60.7 km s$^{\rm -1}$ with dispersion
values of 8.5$-$11.7 km s$^{\rm -1}$, which are consistent with the 
measurements presented here.

\section{Data Analysis}

\subsection{Stellar Atmosphere Parameters}

Initial values for T$_{\rm eff}$ and surface gravity (log(g)) were determined
using the 2MASS photometry given in Table~\ref{tab:basic_params}, the
differential reddening map provided by \citet{ContrerasPena13}, the 
\citet{Gonzalez09} color-temperature relation, a distance of 9.5 kpc
\citep{ContrerasPena18}, [Fe/H] = $-$1.2 \citep{ContrerasPena18}, and the 
K$_{\rm S}$-band bolometric correction relation from \citet{Buzzoni10}. 
However, the final T$_{\rm eff}$ and log(g) values were set by enforcing 
excitation equilibrium with Fe{\sevensize I} and ionization equilibrium 
with Fe{\sevensize I} and Fe{\sevensize II}.  Microturbulence 
($\xi$$_{\rm mic.}$) was determined by removing trends in plots of 
log $\epsilon$(Fe{\sevensize I}) versus line strength.  Additionally, the 
model metallicity was set to the mean of [Fe{\sevensize I}/H] and 
[Fe{\sevensize II}/H].  For each iteration, we interpolated within the ATLAS9
grid\footnote{The grid is available at: 
http://wwwuser.oats.inaf.it/castelli/grids.html.} of $\alpha$-enhanced model 
atmospheres \citep{Castelli04}, and all four parameters were modified 
simultaneously until a satisfactory solution was achieved.  Convergence failed 
for 6/41 members (see Table~\ref{tab:basic_params}) and these objects were 
omitted from further analysis.  The adopted model atmosphere parameters for all
other stars, except for the lone non-member, are provided in 
Table~\ref{tab:basic_params}.

A comparison of the adopted spectroscopic temperatures and those derived via 
photometry, assuming the recommended mean E(B-V) = 0.57 from 
\citet{ContrerasPena13}, indicated that the photometric temperatures were 
$\sim$185 K warmer ($\sigma$ = 89 K), on average.  Agreement between 
the spectroscopic and photometric temperatures was found when the mean E(B-V) 
was set at 0.44, which is within the range of values estimated by previous 
authors \citep[see the discussion in Section 3.1 of][]{ContrerasPena13}.  The 
85 K dispersion between the spectroscopic and photometric T$_{\rm eff}$ 
estimates is also equivalent to the 94 K calibration uncertainty reported by 
\citet{Gonzalez09} for J-K$_{\rm S}$.

\subsection{Equivalent Width and Spectrum Synthesis Analyses}

The equivalent widths (EWs) and abundances of Fe{\sevensize I}, 
Fe{\sevensize II}, Na{\sevensize I}, Si{\sevensize I}, Ca{\sevensize I}, 
Cr{\sevensize I}, and Ni{\sevensize I} were determined using a combination of 
the Gaussian line profile deblender developed for \citet{Johnson14} and the 
\emph{abfind} driver of the local thermodynamic equilibrium (LTE) line analysis
code MOOG\footnote{The MOOG source code is available at: 
https://www.as.utexas.edu/$\sim$chris/moog.html.} \citep[2014 
version;][]{Sneden73}.  The line list, atomic parameters, and reference solar 
abundances were the same as those provided in Table 2 of 
\citet{Johnson15_6273}.  In general, we omitted lines with 
log($\frac{EW}{\lambda}$) $\ga$ $-$4.5 in order to avoid strongly saturated 
features.  A list of the final abundances derived from the EW analyses is 
provided in Table~\ref{tab:abundances}.

For the other elements presented in Table~\ref{tab:abundances}, namely
O{\sevensize I}, Mg{\sevensize I}, Al{\sevensize I}, La{\sevensize II}, and
Eu{\sevensize II}, various issues precluded the use of EW measurements and 
we opted to derive abundances via spectrum synthesis.  Specifically, we used
the MOOG \emph{synth} driver along with line lists from \citet{Johnson14} for
O, \citet{Johnson15_6273} for Mg and Al, \citet{Lawler01_La} for La, and 
\citet{Lawler01_Eu} for Eu.  The reference solar abundances are the same as 
those in \citet{Johnson14} for O and \citet{Johnson15_6273} for the remaining
elements derived from spectrum synthesis.

Since O abundances are sensitive to the total CNO composition, we determined
log $\epsilon$({O{\sevensize I}}) using an iterative process of fitting the 
6300 \AA\ [O{\sevensize I}] feature and several CN bands throughout the 
spectra.  In all syntheses, we held the [C/Fe] ratio fixed at $-$0.3 dex and 
varied the [O/Fe] and [N/Fe] ratios until a satisfactory fit was obtained for
both the O and CN features.  The 6300 \AA\ [O{\sevensize I}] line is also 
blended with Ni{\sevensize I}, which we accounted for by setting the [Ni/Fe] 
ratio to match the value determined from the EW analysis.  The 6696/6698 \AA\ 
Al{\sevensize I} lines are also moderately blended with CN and were only fit 
after determining the relative CNO abundances.

The 6318/6319 \AA\ Mg{\sevensize I} lines used here are relatively weak, 
moderately blended, and can be significantly affected by a shallow but broad 
Ca{\sevensize I} autoionization feature.  The impact of the autoionization 
feature was gauged by examining the continuum suppression between 6316 and 
6320 \AA, and when necessary the strength of the autoionization feature was
modified by changing the Ca abundance in the synthesis.  Although the impact of
the autoionization feature was not too severe in the observed stars, the 
blending and weak line strengths of the 6318/6319 \AA\ triplet limited the 
accuracy of our [Mg/Fe] measurements.  As a result, [Mg/Fe] is likely the most 
poorly constrained measurement presented here.

The La{\sevensize II} and Eu{\sevensize II} lines used here are moderately 
weak (EW $\la$ 50m\AA) but can be affected by hyperfine broadening, isotope
shifts, or both (in the case of Eu).  For La, the 6262 \AA\ line is the 
broadest feature but also the most reliably fit with the \citet{Lawler01_La}
line list.  Therefore, most of the La abundances presented in 
Table~\ref{tab:abundances} rely on this feature.  Similarly, the 6645 \AA\ 
line was the most reliably available feature for Eu.  In all syntheses, 
we adopted the solar \iso{151}{Eu} and \iso{153}{Eu} ratios presented in 
\citet{Lawler01_Eu}.

\subsection{Abundance Uncertainties}

\begin{figure}
\includegraphics[width=\columnwidth]{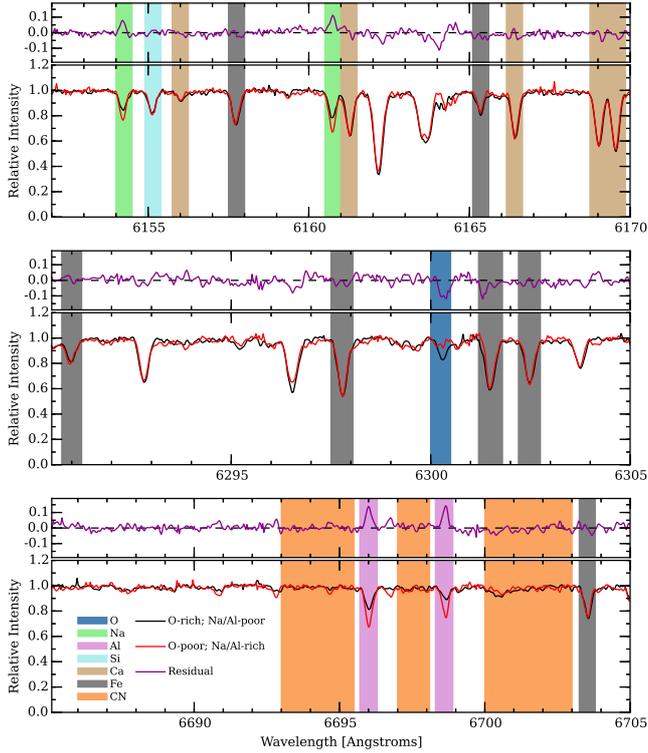}
\caption{Two sample M2FS spectra are shown illustrating the typical data
quality and line strength variations of several light elements.  The black
(2M17374273$-$0315238) and red (2M17373815$-$0314496) spectra represent stars 
with similar T$_{\rm eff}$, log(g), and [Fe/H] but very different light element
compositions.  The shaded regions indicate lines of O, Na, Al, Si, Ca, Fe, and
molecular CN.  Note that the variable line strengths of the V/Ti blend near
6296.5 \AA\ are due to the merging of two spectrograph orders.  Residual plots
are shown in dark magenta above each set of sample spectra.}
\label{fig:f3}
\end{figure}

As noted in Section 2 and demonstrated in Fig.~\ref{fig:f3}, the high S/N of
the M2FS data used in this project suggest that the absolute abundance 
uncertainties are dominated by deficiencies in how radiative transfer is 
calculated and model atmospheres constructed rather than pure measurement 
errors.  However, since the stars utilized in this study span a small range in 
temperature, surface gravity, and metallicity, the relative uncertainty due to 
changes in the model atmosphere parameters is likely the dominant error 
source.  

We determined the abundance uncertainties relative to changes in the model 
atmosphere parameters following the methods outlined in \citet{Johnson18_6569}.
Specifically, we individually adjusted T$_{\rm eff}$, log(g), [Fe/H], and 
$\xi$$_{\rm mic.}$ by 85 K, 0.15 cgs, 0.10 dex, and 0.10 km s$^{\rm -1}$, 
respectively, while holding the other parameters fixed and redetermined the 
abundances of each element.  The $\Delta$T$_{\rm eff}$ value was chosen based 
on the standard deviation of the differences between spectroscopic and 
photometric temperatures (see Section 4.1) while the $\Delta$log(g) uncertainty
was estimated as the mean dispersion in the spectroscopic surface gravity when 
stars were partitioned by T$_{\rm eff}$ into 100 K groups.  Similarly, the 
$\Delta$[Fe/H] value was based on the mean line-to-line [Fe/H] dispersion and 
the $\Delta$$\xi$$_{\rm mic.}$ value was estimated from the 
log $\epsilon$(Fe{\sevensize I}) versus line strength plots.

Abundance differences were calculated between the values listed in
Table~\ref{tab:abundances} and those derived from changes in the model 
atmosphere parameters.  These values were added in quadrature, along with the
error of the mean for each species in order to roughly account for log(gf)
uncertainties, and are reported in Table~\ref{tab:errors}.  Note that for 
cases where only one line was available, we adopted a common fitting 
error of 0.05 dex.  For elements listed in Table~\ref{tab:errors} that are
normalized relative to Fe{\sevensize I}, or Fe{\sevensize II} for species in
the dominant ionization state, the correlated changes in log $\epsilon$(X)
and log $\epsilon$(Fe{\sevensize I}) or log $\epsilon$(Fe{\sevensize II}) are
taken into account.

Additional sources of uncertainty, such as departures from LTE, 1D versus 3D
model atmosphere effects, spherical versus plane parallel calculations, and
modifications to the model atmosphere structure as a result of changes in 
detailed composition, were not taken into account.  Many of these 
effects likely cancel out due to the relative nature of the present analysis,
which was carried out differentially with respect to the metal-poor giant
Arcturus, and the small temperature, gravity, and metallicity range of the 
target stars.  One systematic issue that could affect a portion of our sample 
is the influence of He enhancement on stellar structure and line formation.  
As will be described in Section 5, about 1/4 of our sample may be significantly 
He-enriched.  \citet{Boehm-Vitense79} showed that in cool giants an increase
in He abundance is most likely to affect (strengthen) the neutral atomic
lines of elements with low first ionization potentials, such as Na or Ca.
However, several authors \citep[e.g.,][]{Piotto05,Johnson09,Cunha10} have 
demonstrated that He abundances as high as Y $\sim$ 0.4 likely have little 
effect on the derived [X/Fe] ratios for most elements.

\section{Results and Discussion}

\subsection{Composition Overview}

Previous [Fe/H] estimates for NGC 6402 are based on photometric analyses of 
cluster CMDs \citep[e.g.,][]{ContrerasPena13,Nataf13,ContrerasPena18} and 
find [Fe/H] $\sim$ $-$1.1 to $-$1.4 dex.  The spectroscopic results presented
in Table~\ref{tab:abundances} are in agreement with past work as we find 
a mean [Fe/H] = $-$1.13 dex ($\sigma$ = 0.05 dex), based on a sample of 35
stars.  The small [Fe/H] dispersion of 0.05 dex indicates that, despite being 
among the most massive and luminous clusters in the Galaxy, NGC 6402 does not 
possess a significant intrinsic metallicity spread.

\begin{figure}
\includegraphics[width=\columnwidth]{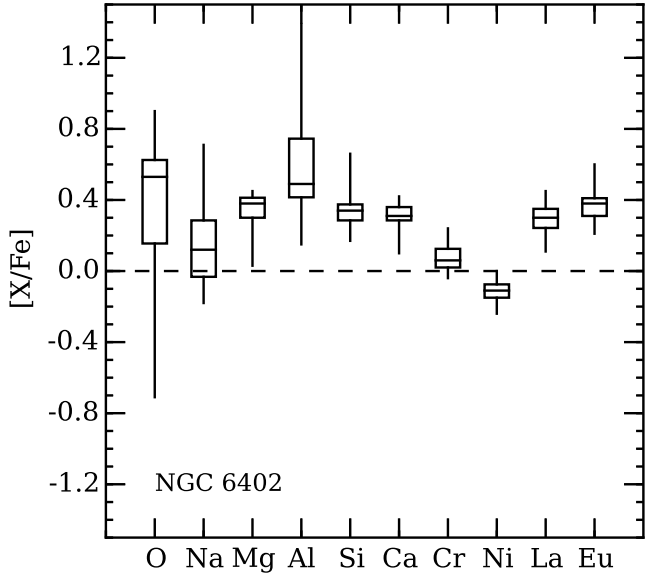}
\caption{A box plot is shown illustrating the [X/Fe] distributions of all
elements analyzed here.  The top, middle, and bottom horizontal lines for each
box illustrate the first, second (median), and third quartiles of the [X/Fe]
distribution.  The vertical lines indicate the minimum and maximum [X/Fe]
ratios.}
\label{fig:f4}
\end{figure}

The overall composition characteristics of NGC 6402 are summarized with the
box plot shown in Fig.~\ref{fig:f4}.  The figure clearly demonstrates that the
light element ratios [O/Fe], [Na/Fe], and [Al/Fe] exhibit significant 
star-to-star variations.  In fact, the full abundance ranges of [O/Fe], 
[Na/Fe], and [Al/Fe] are 1.61, 0.89, and 1.24 dex, respectively.  The clear
line strength changes displayed in Fig.~\ref{fig:f3} indicate that the
light element abundance variations are real and not an artifact of the 
analysis methods.  Smaller but real variations seem to be present for at least 
[Mg/Fe] and [Si/Fe] as well.  Despite these moderate variations, the cluster 
is generally $\alpha$-enhanced with $\langle$[Mg/Fe]$\rangle$ = $+$0.34 dex 
($\sigma$ = 0.10 dex), $\langle$[Si/Fe]$\rangle$ = $+$0.34 dex 
($\sigma$ = 0.10 dex), and $\langle$[Ca/Fe]$\rangle$ = $+$0.31 dex 
($\sigma$ = 0.07 dex).

Unlike the lighter elements, the Fe-peak elements Cr and Ni exhibit 
approximately solar ratios with $\langle$[Cr/Fe]$\rangle$ = $+$0.07 dex
($\sigma$ = 0.07 dex) and $\langle$[Ni/Fe]$\rangle$ = $-$0.11 dex 
($\sigma$ = 0.06 dex).  The heavier neutron-capture elements La and Eu
both exhibit enhanced ratios with $\langle$[La/Fe]$\rangle$ = $+$0.29 dex 
($\sigma$ = 0.08 dex) and $\langle$[Eu/Fe]$\rangle$ = $+$0.37 dex 
($\sigma$ = 0.09 dex).  The moderately low mean [La/Eu] = $-$0.08 dex and 
0.1 dex dispersion suggest that the heavy element pollution within NGC 6402
was well-mixed and did not contain significant enrichment from the slow
neutron-capture process.

A comparison of the mean abundance distributions shown in Fig.~\ref{fig:f4}
with the typical composition characteristics found in other inner Galaxy
globular clusters \citep[e.g., see Fig. 9-10 of][]{Johnson18_6569} indicates
that NGC 6402 follows the common trend of [$\alpha$/Fe] $\sim$ $+$0.3 dex,
[Fe-peak/Fe] $\sim$ $+$0.0 dex, and [La/Eu] $<$ 0.0 dex.  The heavier $\alpha$,
Fe-peak, and neutron-capture element data are also consistent with bulge and
disk field stars of similar [Fe/H] found within the inner Galaxy 
\citep[e.g.,][]{Johnson14,Bensby17}.  This suggest that the primordial gas 
from which NGC 6402 formed possessed a composition similar to the Milky Way 
field populations near [Fe/H] $\sim$ $-$1.  However, a more detailed 
examination of NGC 6402's light element composition reveals a particularly 
interesting formation history.

\subsection{Identifying Multiple Populations via Light Element Variations}

As mentioned in Section 1, globular clusters are generally composed of two or
more populations with different light element compositions.  
However, detailed comparisons of sub-populations within different systems are
difficult to interpret because no specific set of chemical properties define
the various cluster components.  Furthermore, nomenclature varies among 
authors with little agreement existing beyond the notion that ``first" 
generation/population stars have compositions similar to halo field stars and
``second" generation/population stars exhibit higher He, N, and Na with lower
C and O abundances.  For the purposes of this work we have adopted the
nomenclature of \citet{Carretta15}, which identified at least five independent
groups in NGC 2808 associated with primordial (``halo-like": P1/P2), 
intermediate (diluted: I1/I2), and extreme (pure polluted material: E)
compositions.  We also adopt the interpretation presented by \citet{Dantona07},
\citet{Milone15_n2808}, and \citet{Dantona16} that the P1 group is the sole 
primordial population, the P2, I1, and I2 stars represent diluted second 
generation compositions, and that the E second generation stars trace a 
combination of heavy pollution and possibly some \emph{in situ} mixing.  Note 
that the P1, P2/I1, I2, and E groups may be equivalent to the B, C, D, and E 
populations identified in \citet{Milone15_n2808}, respectively.

\begin{figure}
\includegraphics[width=\columnwidth]{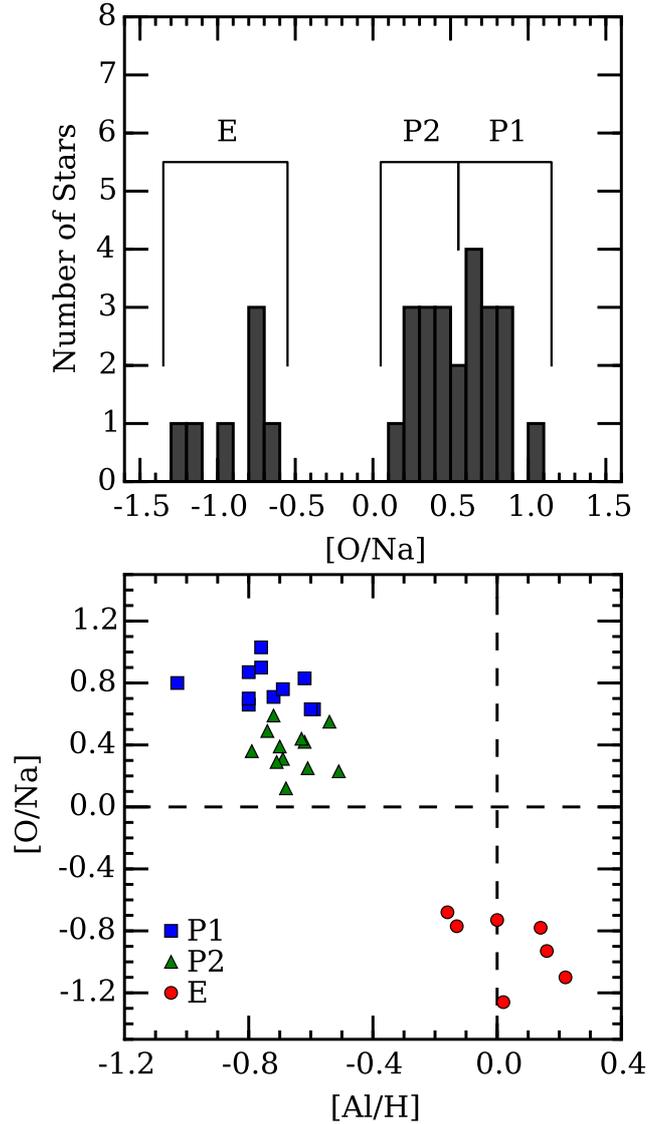}
\caption{\emph{Top:} a histogram of [O/Na] ratios for NGC 6402 indicates the 
chemical distinction between P1, P2, and E population stars.  The data are 
sampled using 0.1 dex bins.  Note the paucity of stars with $-$0.6 $<$ [O/Na] 
$<$ $+$0.1 that match the typical I1 and I2 groups.  \emph{Bottom:} [O/Na] is 
plotted as a function of [Al/H] to illustrate the composition differences
between P1, P2, and E stars.}
\label{fig:f5}
\end{figure}

As illustrated in Fig.~\ref{fig:f5}, the NGC 6402 data were partitioned into 
sub-populations using each star's [O/Na] ratio\footnote{For cases in which 
[O/Fe] or [Na/Fe] could not be measured, identification as a P1, P2, or E star 
was based on [Al/Fe].}.  The bottom panel of Fig.~\ref{fig:f5} further 
indicates that the P1$+$P2 and E groups are easily distinguished using any 
combination of O, Na, and Al abundances, but the P1/P2 separation is more 
ambiguous.  The [O/Na] dispersion for stars with [O/Na] $>$ 0.0 dex is 0.24 
dex, which is $\sim$30 per cent larger than the mean [O/Na] measurement error 
estimated from the individual [O/Fe] and [Na/Fe] values provided in 
Table~\ref{tab:errors}.  The broad and potentially bimodal [O/Na] distribution 
shown in the top panel of Fig.~\ref{fig:f5} suggests either that two low-Na
populations are present or that the P1$+$P2 group is not homogeneous in 
composition.  

\begin{figure}
\includegraphics[width=\columnwidth]{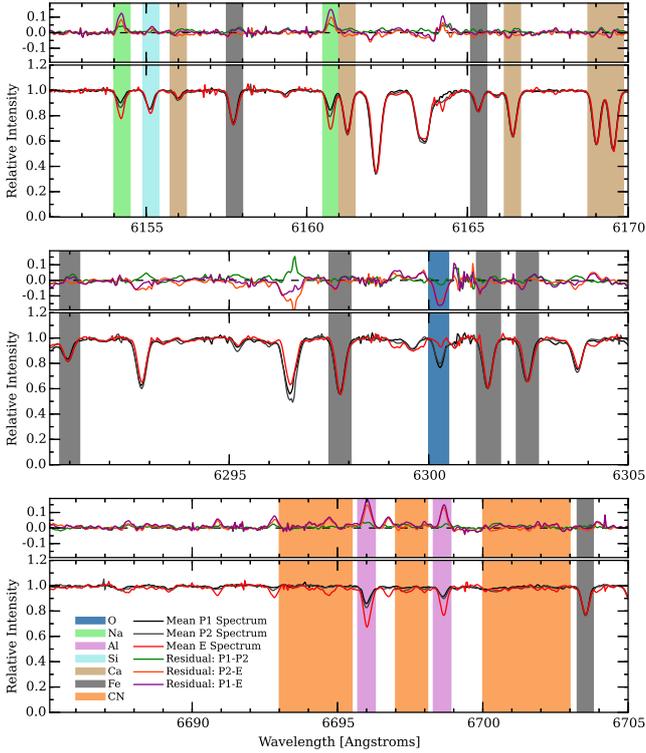}
\caption{Similar to Fig.~\ref{fig:f3}, mean combined spectra for stars in the
P1, P2, and E groups with T$_{\rm eff}$ $\sim$ 4300 K ($\sim$6 stars in each
group) are compared.  Line depth residuals for the P1$-$P2, P2$-$E, and P1$-$E 
mean spectrum pairs are illustrated with the green, dark orange, and dark 
magenta lines, respectively.  Note the significant variations in O, Na, Al, and
CN line depths between especially the P1 and E groups.}
\label{fig:f6}
\end{figure}

Photometric evidence from \emph{HST} ``chromosome" maps suggests that intrinsic
composition variations may be present among first generation stars in some 
clusters.  However, no clear correlation exists between the observed color 
dispersions and variations in [O/Fe] or [Na/Fe] for first generation stars, 
which suggests that the color extensions may only reflect changes in He and/or 
N \citep{Milone17_atlas,Milone18_1gspread,Lardo18}.  Since the P1 and P2 groups 
identified here exhibit nearly identical O-Na distributions as those presented 
in \citet{Carretta15} for NGC 2808 (see also Section 5.4) and 
\citet{Milone15_n2808} find the NGC 2808 P1 and P2 stars to exhibit very 
different near-UV/optical colors, we conclude that the NGC 6402 stars with 
[O/Na] $>$ 0.0 dex form two distinct populations.  Further evidence supporting 
clear composition differences between the P1 and P2 stars can be seen in the 
combined spectra of Fig.~\ref{fig:f6} where the P2 population exhibits weaker 
O and stronger Na/Al lines.

Table~\ref{tab:populations} outlines the mean chemical properties of the P1,
P2, and E groups and shows that many of the composition differences are of high
statistical significance.  As was already evident from Fig.~\ref{fig:f5},
results from a Welch's t-test analysis indicate that significant differences 
exist between the various NGC 6402 populations when considering [O/Fe], 
[Na/Fe], and [Al/Fe].  However, the P1 and P2 groups exhibit similar [Mg/Fe]
and [Si/Fe] abundances while the E population has significantly lower [Mg/Fe]
and higher [Si/Fe].  Similarly, the E group has higher [Ca/Fe] than the P1 
population, but the data are inconclusive regarding whether the P2 [Ca/Fe] 
abundances differ from the P1 and/or E populations.  For elements heavier than 
Ca, Table~\ref{tab:populations} indicates that the various sub-populations are 
chemically indistinguishable (see also Fig.~\ref{fig:f6}).

\subsubsection{Population Ratios}

\begin{figure}
\includegraphics[width=\columnwidth]{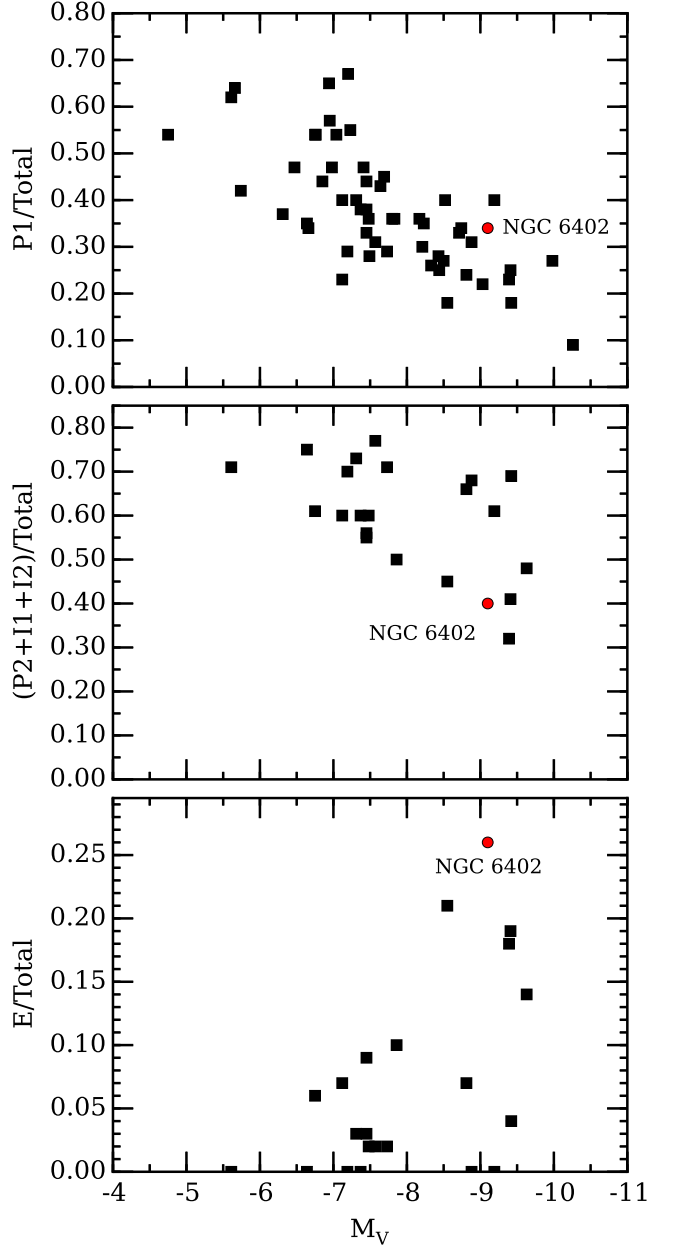}
\caption{The observed ratios of first generation (P1; top panel), intermediate
second generation (P2, I1, and I2; middle panel), and extreme second generation
(E; bottom panel) stars relative to the cluster totals are plotted as a 
function of each cluster's absolute magnitude (a tracer for mass).  Data are 
shown for various literature sources (filled black boxes) and NGC 6402 (filled 
red circles).  The literature data for first generation stars are from
\citet{Milone17_atlas} while the intermediate and extreme second generation 
ratio data are from \citet{Carretta09_gir}.}
\label{fig:f7}
\end{figure}

Recent large sample analyses by \citet{Carretta09_gir} and 
\citet{Milone17_atlas} revealed several global trends: (1) globular 
clusters are numerically dominated by intermediate compositions stars; (2)
most clusters have first generation fractions of $\sim$30$-$40 per cent; (3)
the fraction of first generation stars strongly declines as a function of 
increasing present-day mass; and (4) extreme second generation stars with very
low [O/Fe] are rare, predominantly located in the most massive clusters with
extended blue HBs, and when found are present at only the 5$-$20 per cent 
level.  Fig.~\ref{fig:f7} summarizes these observations and shows that the 
intermediate composition ratio generally declines with increasing cluster mass 
as well.

For NGC 6402, we find the P1, P2, and E groups to represent 34 per cent 
(12/35), 40 per cent (14/35), and 26 per cent (9/35) of our sample,
respectively.  Fig.~\ref{fig:f7} indicates that these population ratios are 
in general agreement with the bulk trend observed for other Galactic globular
clusters.  Interestingly, the extreme population fraction in NGC 6402 may be
among the highest of any known cluster.  Although the presence of extreme 
composition stars is not surprising given the cluster's extended blue HB, we 
note that similar mass clusters generally contain large populations of I1 and 
I2 stars as well.  However, Fig.~\ref{fig:f5} indicates that the I1/I2 
intermediate composition stars may be entirely absent from NGC 6402.  As a 
result the P2 fraction may be unusually high in NGC 6402, especially when 
compared against the NGC 2808 P2 component that constitutes only 25 per cent
of the \citet{Carretta15} sample.

\subsection{Light Element (Anti-)correlations}

\begin{figure}
\includegraphics[width=\columnwidth]{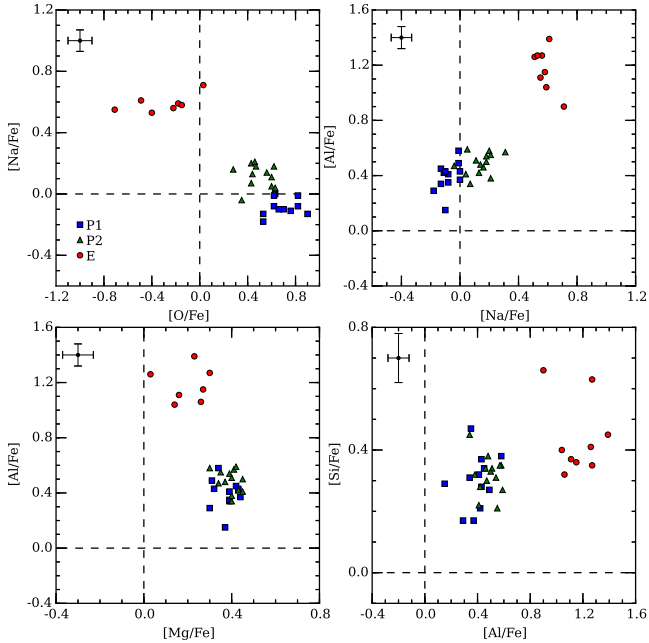}
\caption{Combinations of [O/Fe], [Na/Fe], [Mg/Fe], [Al/Fe], and [Si/Fe] are
plotted for NGC 6402.  In all panels the filled blue boxes, green triangles,
and red circles indicate stars belonging to the P1, P2, and E populations,
respectively, following the notation from \citet{Carretta15}.  Note the large
abundance gap present for all element combinations.}
\label{fig:f8}
\end{figure}

Fig.~\ref{fig:f8} shows that NGC 6402 exhibits the same light element
(anti-)correlations that are present in nearly all old massive globular
clusters.  Similar to other clusters with extended blue HBs, the 
(anti-)correlated abundance patterns in NGC 6402 are present not only for 
[O/Fe] and [Na/Fe] but extend to [Mg/Fe], [Al/Fe], and [Si/Fe] as well.
However, (anti-)correlations involving [Mg/Fe] and [Si/Fe] are driven solely
by the E population.  As established in Section 5.2 and reinforced through 
Fig.~\ref{fig:f8}, the P1 and P2 populations exhibit noticeably different 
[O/Fe], [Na/Fe], and [Al/Fe] abundances but indistinguishable [Mg/Fe] and 
[Si/Fe].  

An examination of Fig. 2 in \citet{Prantzos07} indicates that, depending on the
He abundance of the P2 group, temperatures as low as 45$-$60 MK could be 
sufficient to modify P1 composition gas such that [O/Fe] declines, [Na/Fe] and 
[Al/Fe] increase, and [Mg/Fe] and [Si/Fe] remain unchanged.  If burned at 
45$-$60 MK, the \iso{25}{Mg} and \iso{26}{Mg} isotopes could be particularly 
powerful temperature discriminators.  However, previous work on other clusters 
has shown that clear variations in the \iso{24}{Mg}, \iso{25}{Mg}, and 
\iso{26}{Mg} isotope ratios are only detected when [Al/Fe] enhancements are 
large \citep[e.g.,][]{Shetrone96,Yong03,Sneden04,DaCosta13,Ventura18}.  
Although the P2 gas could have been burned at higher temperatures and then 
diluted with ``pristine" P1 composition material, no single dilution 
model fits through the P1, P2, and E populations shown in Fig.~\ref{fig:f8}
(see also Section 5.4).  Therefore, we conclude that the P2 group is not a 
result of simple dilution between P1 and E composition gas and instead 
represents an independent pollution event.

\begin{figure}
\includegraphics[width=\columnwidth]{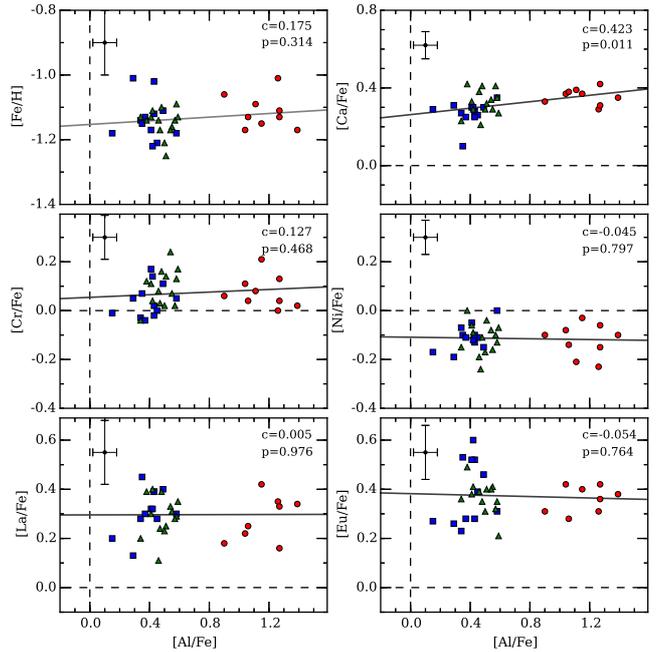}
\caption{Similar to Fig.~\ref{fig:f8}, [Fe/H], [Ca/Fe], [Cr/Fe], [Ni/Fe],
[La/Fe], and [Eu/Fe] are plotted as a function of [Al/Fe].  The solid grey 
lines show the best-fit linear functions through each distribution.  The
numerical ``c" and ``p" values provided in each panel are the Pearson
correlation coefficients and associated statistical p-values from the fits.
Note that only [Ca/Fe] shows any evidence supporting a statistically
significant correlation with [Al/Fe].}
\label{fig:f9}
\end{figure}

As mentioned above, the Mg-Al anti-correlation and Al-Si correlation shown in
Fig.~\ref{fig:f8} are driven by composition differences between the P1$+$P2 and
E populations.  Nucleosynthesis studies have shown that significant changes to
the [Mg/Fe] and [Si/Fe] abundances can only occur when burning temperatures 
exceed $\sim$65 MK \citep[e.g.,][]{Karakas03,Prantzos07,Prantzos17,
Ventura18}.  These results therefore place a lower limit of about 65 MK on the 
burning conditions responsible for producing the gas from which the E 
population formed, and indicate further that the P1, P2, and E groups in 
NGC 6402 did not originate from a single pollution source (see also Section 
5.4).  

Interestingly, Fig.~\ref{fig:f9} shows that the correlated abundance variations
may reach to elements as heavy as Ca since a potentially significant 
correlation between [Al/Fe] and [Ca/Fe] is detected.  Furthermore, the Welch's 
t-test results summarized in Table~\ref{tab:populations} indicate with high 
confidence that the mean [Ca/Fe] abundance is higher in the E population than 
the P1 group.  A marginal detection of the P2 population having a higher mean 
[Ca/Fe] abundance than the P1 group is also observed; however, we consider this
result to be spurious since neither [Mg/Fe] nor [Si/Fe] vary significantly 
between the P1 and P2 stars.  If the Al-Ca correlation is confirmed, NGC 6402
would join clusters such as NGC 2808 \citep{Carretta15,Mucciarelli15} and 
NGC 2419 \citep{Cohen12,Mucciarelli12} in possessing light element correlations
that extend to elements heavier than Si.  Such a confirmation would indicate
that the gas from which the E population stars in NGC 6402 formed was
processed at temperatures $>$100 MK \citep{Ventura12,Iliadis16}.

However, we stress two important caveats related to the Al-Ca correlation
detection: (1) Table~\ref{tab:populations} shows that the mean difference in 
[Ca/Fe] between the P1 and E groups is only 0.09 dex and therefore noticeably
smaller than the P1-E [X/Fe] differences involving lighter elements; and (2) 
the Al/Ca-rich population is likely He-enhanced, and as mentioned in Section 
4.3 neutral Ca is one of the species that may be susceptible to line strength 
enhancements due to changes in the He mass fraction.  Stronger confirmation of 
very high temperature burning in NGC 6402 may be obtained by searching for 
similar correlations involving Sc and/or K.  These elements have been shown to 
vary in at least NGC 2808 and NGC 2419 \citep{Cohen12,Mucciarelli12,Carretta15,
Mucciarelli17} and possess stronger correlation signatures owing to their 
intrinsically lower abundances.

\subsubsection{Missing Intermediate Composition Stars}

The most notable observation from Fig.~\ref{fig:f8} is the paucity of 
intermediate composition (I1/I2) stars having [O/Fe] $\sim$ $+$ 0.1 dex, 
[Na/Fe] $\sim$ $+$0.3 dex, and [Al/Fe] $\sim$ $+$0.8 dex.  The discrete nature 
of the light element (anti-)correlations shown in Fig.~\ref{fig:f8} is not 
surprising, but previous analyses have shown that clusters do not generally
have E composition stars without significant populations of I1/I2 composition
stars \citep[e.g.,][]{Carretta09_gir}.  In fact, the $\omega$ Cen groups with
[Fe/H] $\ga$ $-$1.3 are the only other populations that exhibit a large number
of E composition stars with a simultaneous paucity of I1/I2 stars 
\citep{Johnson10,Marino11_omcen}\footnote{Similar composition gaps have been 
found in a few other clusters \citep[e.g.,][]{Carretta14}, but these cases tend
to have sample sizes of $\sim$10 stars or fewer.  Clusters such as M 4 
\citep{Marino08,Villanova11} also exhibit significant composition gaps but do
not contain E stars.}.  However, we note that a similar composition
gap could also be present in NGC 2419 based on analyses of the cluster's HB 
morphology \citep{DiCriscienzo11} and the UVES sample of \citet{Mucciarelli15}.

\begin{figure}
\includegraphics[width=\columnwidth]{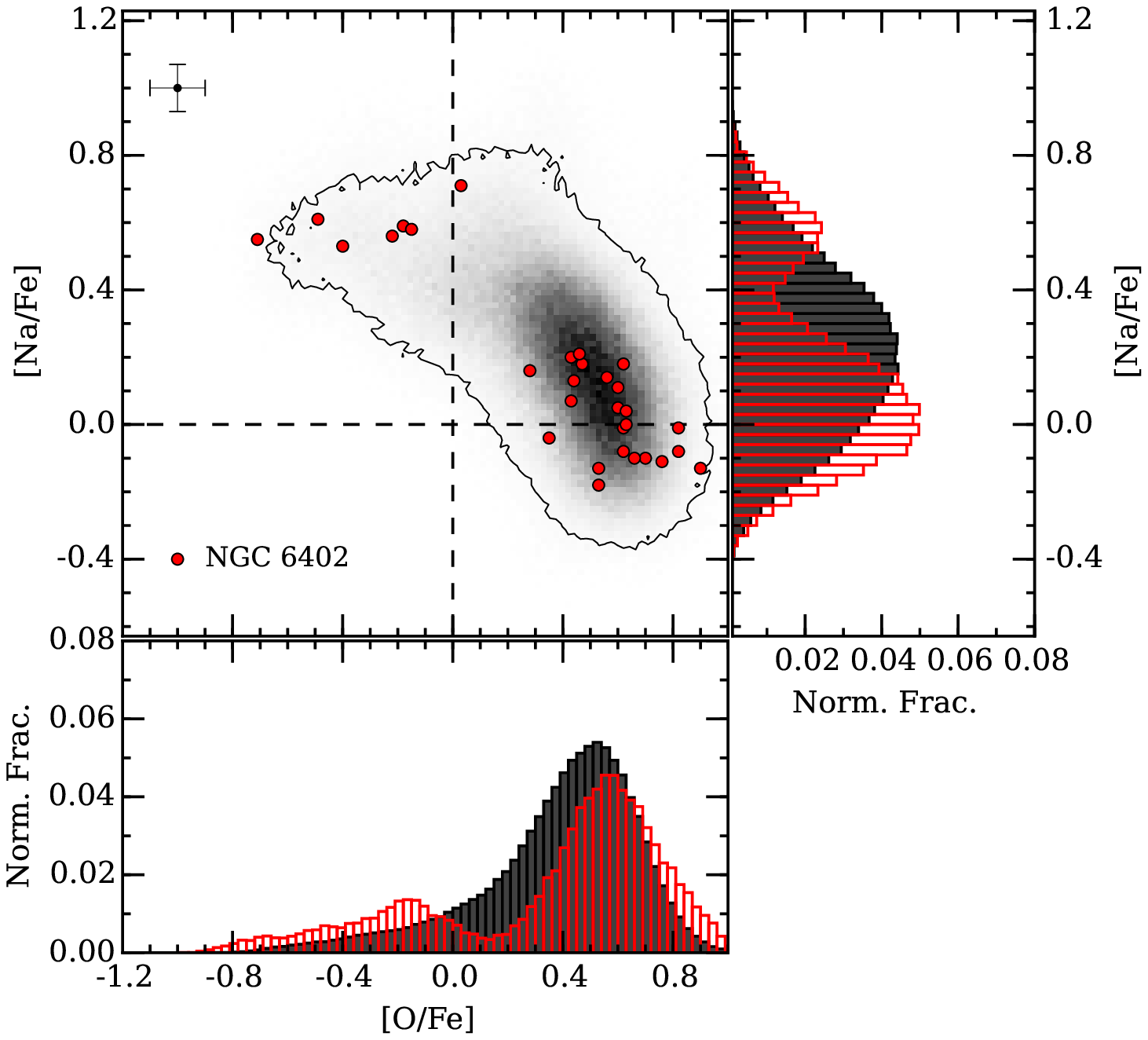}
\caption{The [Na/Fe] abundances for NGC 6402 (filled red circles) are plotted
as a function of [O/Fe] and compared against similar clusters (density plot)
with $-$1.3 $<$ [Fe/H] $<$ $-$0.8 from \citet{Carretta09_gir,Carretta09_uves},
\citet{Yong14_6266}, \citet{Carretta15}, \citet{Lapenna15_6266}, and
\citet{Johnson17_6229,Johnson18_6569}.  The literature sources were shifted to
have approximately the same upper [O/Fe] and lower [Na/Fe] bounds as NGC 6402,
and were resampled 10$^{\rm 4}$ times.  Each resampling was convolved with a
Gaussian error distribution of 0.1 dex in [X/Fe].  The contour outlines a
region where 95 per cent of the resampled literature data reside.  Histograms
for the literature (grey) and similarly resampled NGC 6402 data (red) are shown
for both [O/Fe] and [Na/Fe].}
\label{fig:f10}
\end{figure}

The unusual light element composition of NGC 6402 is further emphasized in 
Fig.~\ref{fig:f10} where we compare the cluster's [O/Fe] and [Na/Fe] abundances
against those of other similar metallicity clusters shifted onto the same 
scale\footnote{The literature [O/Fe] and [Na/Fe] data have been shifted by eye
to have approximately the same mean composition as the ``P1" group (see Section
5.2.1) in NGC 6402.  These shifts aim to minimize systematic effects due to the
use of different solar abundances, line lists, analysis methods, and non-LTE
corrections.}.  When convolved with a typical measurement uncertainty of 0.1 
dex and resampled to produce a probability distribution, we find that NGC 6402
exhibits both a very extended O-Na anti-correlation and a paucity of the most
common intermediate composition stars found in clusters with [Fe/H] $\sim$ 
$-$1.  Fig.~\ref{fig:f10} also indicates that even if other clusters actually 
have bimodal O-Na distributions, the mean differences between the two groups 
are not as extreme as in NGC 6402.

An important question to address is whether the composition gaps
shown in Fig.~\ref{fig:f8} and Fig.~\ref{fig:f10} are real or due to 
sampling effects.  Since the O-Na distribution from Fig.~\ref{fig:f8} 
includes 32 stars for NGC 6402, we investigated the influence of sampling by 
performing 10$^{\rm 5}$ random drawings of 32 stars, each with an added random 
Gaussian error distribution ($\sigma$ = 0.1 dex), from two different parent 
populations.  The first population from which samples were drawn followed the
literature distribution shown in Fig.~\ref{fig:f10}, which is relatively smooth
and peaks near the observed gap in NGC 6402.  The second population from which 
samples were drawn traced the clumpy and extended NGC 2808 distribution from 
\citet{Carretta15}.  In each case, we measured the frequency with which the 
random drawings selected at least 7 E composition stars and also failed 
to include any stars within the ([O/Fe],[Na/Fe]) coordinates ($+$0.2,$+$0.8), 
($-$0.35,$+$0.4), ($+$0.1,$+$0.1), and ($+$0.6,$+$0.5).  

For the first distribution, only 0.004 per cent of the simulations produced an 
O-Na gap similar to that observed in NGC 6402.  Similarly, the second 
distribution produced a gap with a frequency of only 0.057 per cent.  We also
found that for a sample size of 32 the expected numbers of stars residing in
the O-Na gap of Fig.~\ref{fig:f10} are 9 and 6 if the parent populations 
trace the continuous and clumpy (NGC 2808) distributions, respectively.  A 
combination of measurement uncertainty and/or stochastic enrichment will likely
produce at least some I1/I2 stars in a larger sample of NGC 6402.  However,
the present data indicate a low probability that NGC 6402's true O-Na 
distribution contains a significant number of I1/I2 stars.  Fig.~\ref{fig:f1}
does not reveal any clear observational bias, and the mean spectra illustrated
in Fig.~\ref{fig:f6} show a substantial change in O and Na line strengths 
between the P2 and E groups.  If I1/I2 stars are present in NGC 6402 then we 
estimate an upper limit of only $\sim$10$-$15 per cent of the total cluster 
population, which is far lower than the $\sim$25 per cent occurrence rate in
NGC 2808 \citep{Carretta15}.

\subsection{Nucleosynthesis Implications from a Comparison with NGC 2808}

\begin{figure*}
\includegraphics[width=\textwidth]{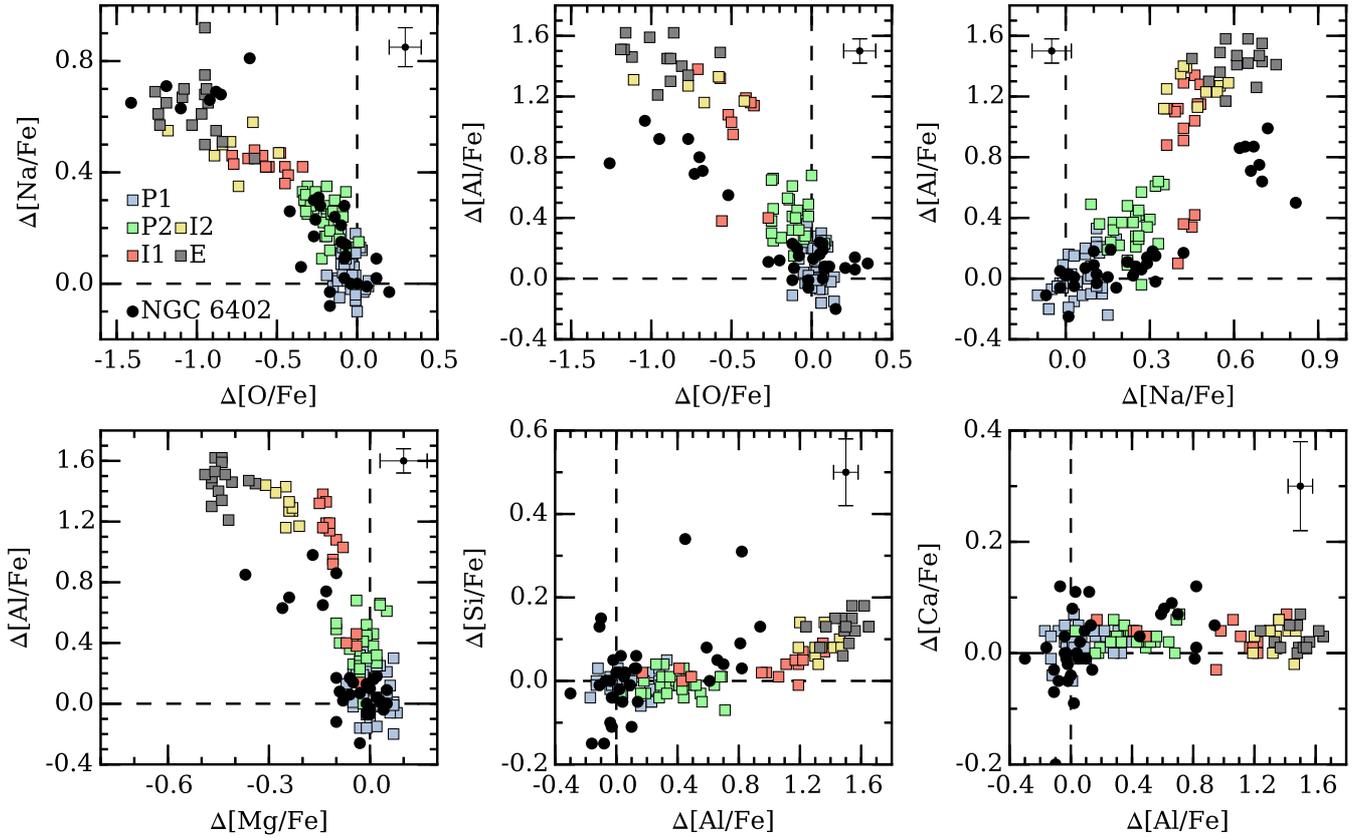}
\caption{NGC 6402 (filled black circles) and NGC 2808 (filled boxes) chemical
abundance data are compared using combinations of $\Delta$[O/Fe],
$\Delta$[Na/Fe], $\Delta$[Mg/Fe], $\Delta$[Al/Fe], $\Delta$[Si/Fe], and 
$\Delta$[Ca/Fe] where $\Delta$[X/Fe] represents the difference between a star's
[X/Fe] ratio and the cluster's assumed primordial composition.  The NGC 2808 
data and population classifications (P1, P2, I1, I2, and E) are from 
\citet{Carretta15} and \citet{Carretta18}.  The NGC 6402 and NGC 2808 data have
been shifted to have the same mean [X/Fe] ratios for the P1 groups.  See text 
for details.}
\label{fig:f11}
\end{figure*}

In Fig.~\ref{fig:f11} we compare the relative abundance changes 
($\Delta$[X/Fe]) in [O/Fe], [Na/Fe], [Mg/Fe], [Al/Fe], [Si/Fe], and [Ca/Fe] 
between the P1 and other sub-populations for NGC 6402 and NGC 2808.  The two
clusters share similar present-day masses, ages, mean [Fe/H] values, extended 
blue HBs, and exhibit spectroscopic evidence of various populations with unique
chemical compositions \citep[e.g.,][]{Carretta15,Carretta18}.  However, their 
detailed composition patterns are not the same.

The $\Delta$[O/Fe]-$\Delta$[Na/Fe] panel in Fig.~\ref{fig:f11} best illustrates
the missing I1/I2 population of NGC 6402 compared to NGC 2808, and shows that
despite the composition gap the relative O-Na distributions for the P1, P2, and
E groups between the two clusters are nearly identical.  However, the NGC 6402
$\Delta$[Mg/Fe], $\Delta$[Al/Fe], and $\Delta$[Si/Fe] values for the E group
are lower than in NGC 2808 by approximately 0.25 dex, 0.7 dex, and 0.05 dex, 
respectively.  To first order, the E population composition differences between
NGC 6402 and NGC 2808 can be understood in terms of burning temperatures.

\begin{figure*}
\includegraphics[width=\textwidth]{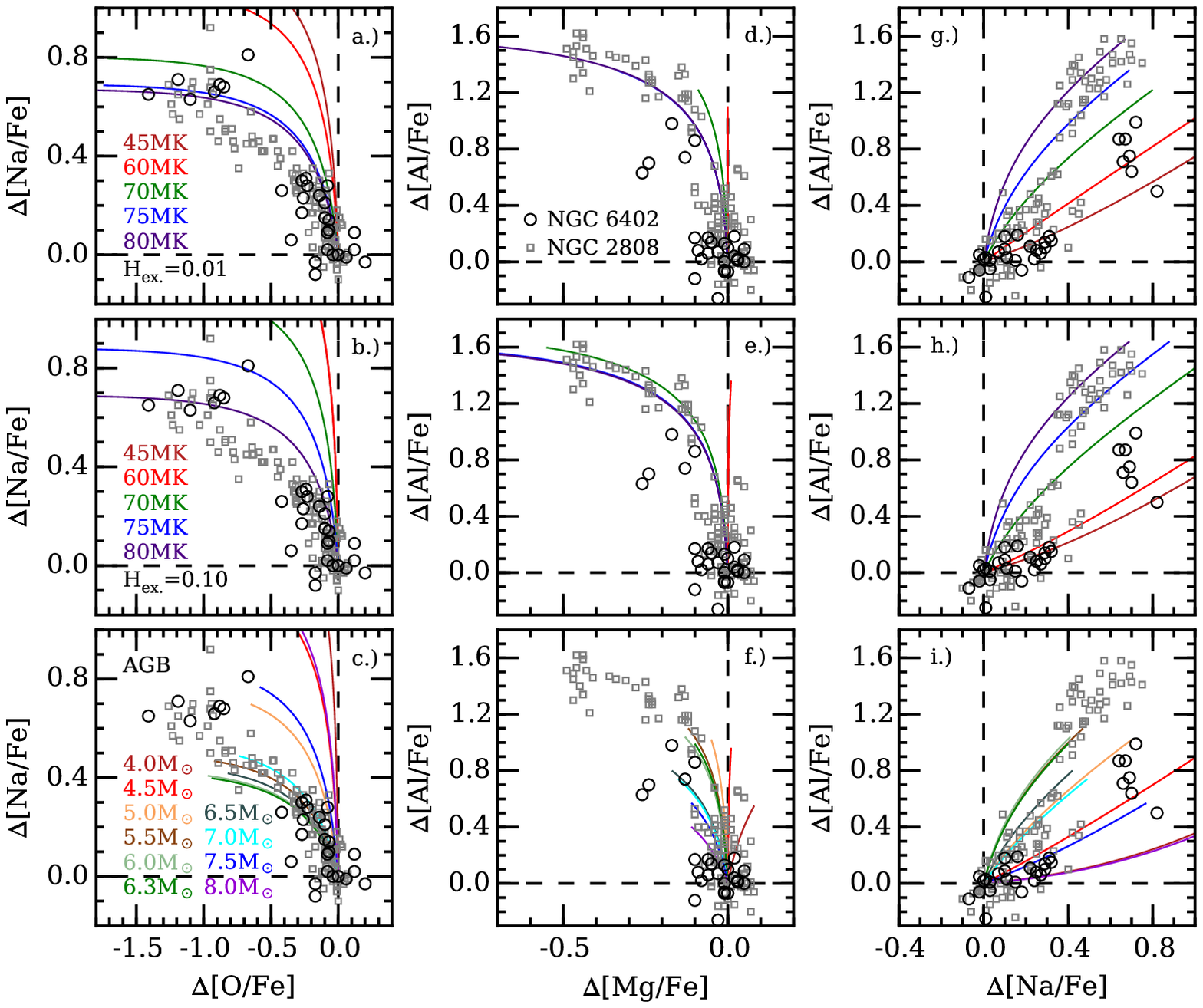}
\caption{Similar to Fig.~\ref{fig:f11}, the relative changes in the O-Na 
(panels a$-$c), Mg-Al (panels d$-$f), and Na-Al (panels g$-$i) 
(anti-)correlations are compared for NGC 6402 (large open circles) and
NGC 2808 (small grey boxes).  The top row (panels a, d, and g) overplots the 
composition changes and resulting dilution curves using data from 
\citet{Prantzos07} when P1 gas is burned at various temperatures of 45$-$80 MK 
and reaches a hydrogen exhaustion (H$_{\rm ex.}$; 
(X$_{\rm O}$$-$X)/X$_{\rm O}$) value of 0.01.  The middle row (panels b, e, and
h) are similar except the hydrogen exhaustion value is an order of magnitude 
higher at 0.10.  The bottom row (panels c, f, and i) are similar except the 
different colored lines represent dilution curves based on the Z = 
10$^{\rm -3}$ AGB ejecta models of \citet{Ventura13} for masses of 4$-$8 
M$_{\sun}$.  The key conclusions of this plot are: (1) Mg-depletion
requires burning temperatures $>$ 65 MK to explain the most extreme stars in 
both clusters; (2) no single dilution curve fits any cluster; (3) the E 
group in NGC 2808 may be understood as a result of hydrostatic burning at 
75$-$80 MK, but the extreme stars in NGC 6402 seem to require $\sim$70 MK 
conditions along with a higher ratio of pristine material to lower the 
[Na/Fe] and [Al/Fe] abundances; (4) the intermediate composition stars of 
NGC 2808 are fit by $\sim$5.5$-$7 M$_{\sun}$ AGB ejecta; and (5) the P2 stars 
of NGC 6402 are not fit by a single set of AGB yields, but may be compatible 
with $\sim$7$-$8 M$_{\sun}$ AGB ejecta if the [O/Fe] (for 7.5$-$8 M$_{\sun}$
models), [Na/Fe] (for 7.5$-$8 M$_{\sun}$ models), and/or [Al/Fe] (for 7 
M$_{\sun}$ models) yields can be reduced.}
\label{fig:f12}
\end{figure*}

For example, Fig.~\ref{fig:f12} utilizes the nucleosynthesis calculations of
\citet{Prantzos07} to show that the E group $\Delta$[O/Fe] and $\Delta$[Na/Fe]
values for both clusters may be well described using burning temperatures of
at least 70$-$80 MK.  However, the significantly larger $\Delta$[Mg/Fe] and
$\Delta$[Al/Fe] values reached by NGC 2808's E population suggest the cluster
was predominantly polluted by gas reaching at least 75$-$80 MK.  In contrast,
the NGC 6402 E population data are better fit with a burning temperature of 
$\sim$70 MK.  A higher burning temperature for the E population in NGC 2808 is 
also consistent with its larger Si enhancements since \iso{28}{Si} production 
is a relatively monotonic function of temperature in the $\sim$65$-$125 MK 
range \citep[e.g.,][]{Prantzos17}.  As a consequence of the very high 
temperatures required to modify elements heavier than Si, the Al-Ca correlation
signature observed in NGC 6402, but not NGC 2808, seems likely to be a spurious
result.

Interestingly, Fig.~\ref{fig:f12} shows that Mg destruction and Na/Al 
production are strong functions of the hydrogen exhaustion (H$_{\rm ex.}$) 
value, which is used as a proxy for burning time or extent.  Although NGC 6402 
and NGC 2808 are largely fit by the 70$-$80 MK dilution curves when 
H$_{\rm ex.}$ = 0.01, such low values may not permit sufficient He production 
to explain the 0.10$-$0.15 He mass fraction ($\Delta$Y) enhancements likely
present in the E populations of each cluster.  Larger H$_{\rm ex.}$ values can
also fit the data, but at least for NGC 6402 such a scenario would require 
additional dilution to decrease the maximum $\Delta$[Na/Fe] and $\Delta$[Al/Fe]
ratios.  We note that \citet{Prantzos17} did include a small initial dilution 
of 4$-$5 per cent pristine material to explain the light element variations in 
NGC 2808.  Small H$_{\rm ex.}$ values may be acceptable if additional gas with 
the proper composition (e.g., He-enhanced) was present in the early cluster 
environment from sources such as Wolf-Rayet stars \citep[e.g.,][]{Smith06} or
Luminous Blue Variables \citep[e.g.,][]{Dufour97,Smith04}.  Such a scenario 
could boost the E population's He abundance to higher levels than would be 
expected based on the abundances of other light elements alone.  Temperatures 
exceeding 80 MK also lower the Na and Al yields but may destroy too much Mg and
produce too much Si to fit the observations \citep[e.g., see 
Fig. 3 of][]{Prantzos17}.

Further inspection of Fig.~\ref{fig:f11} and Fig.~\ref{fig:f12} indicates 
that [Al/Fe] abundance differences between NGC 6402 and NGC 2808 extend to the
P2 populations as well.  None of the nucleosynthesis dilution 
curves in Fig.~\ref{fig:f12} pass through both the P2 and E groups of either
cluster, and the $\Delta$[Na/Fe]-$\Delta$[Al/Fe] plots in particular seem to
rule out that the P2 and E stars could have originated from the same pollution
sources.  These observations reinforce the conclusion by \citet{Carretta18}
that at least NGC 2808 cannot be fit by a simple dilution model, and we 
conclude that at least two different enrichment sources are required to 
explain the NGC 2808 and NGC 6402 data.  Although the pure hydrostatic burning
curves provide an unsatisfactory explanation for the P2 composition stars in 
both clusters, the \citet{Ventura13} AGB models shown in the bottom rows of 
Fig.~\ref{fig:f12} are a better fit.  

AGB models struggle to reproduce the abundance patterns of the E populations in
most clusters\footnote{Strong Mg depletion is a particular problem, especially 
at higher metallicities \citep[e.g.,][]{Dantona16}, but may be remedied with 
updated models \citep{DiCriscienzo18}.}, but are capable of explaining a 
majority of the intermediate composition stars.  Fig.~\ref{fig:f12} shows that 
the NGC 2808 P2, I1, and I2 stars are well-fit by the products of 5.5$-$7 
M$_{\sun}$ AGB ejecta, but the NGC 6402 P2 stars are not completely fit by any 
single model set.  However, the NGC 6402 P2 stars would be consistent with
the dilution curves from the 7$-$8 M$_{\sun}$ AGB models if the O and/or Na
yields from the 7.5 M$_{\sun}$ AGB model or the Al yield from the 7 M$_{\sun}$
AGB model could be reduced.  The new calculations presented in 
\citet{DiCriscienzo18} provide a decrease in [Na/Fe] for the 7$-$7.5 M$_{\sun}$
models, but the substantial increase in [Al/Fe] precludes these calculations
from improving the fits in Fig.~\ref{fig:f12}.  Regardless of the modeling 
difficulty, we conclude that the P2 composition stars in NGC 6402 were polluted
by a different class or (higher?) mass range of objects than those in NGC 2808.  

\subsection{Impact of Light Element Variations on Post-RGB Evolution}

Several authors have demonstrated a close connection between the magnitude of a
globular cluster's light element variations and the color/temperature extent
of its HB \citep[e.g.,][]{Carretta07,Gratton10,Milone14}.  Although the 
detailed morphology of a cluster's HB is a complicated mixture of age, 
metallicity, He abundance, rotation, RGB mass loss, and several atmospheric
effects \citep[e.g.,][]{Catelan09}, some additional insight may be gained by
examining cluster sets that minimize differences between these parameters.  A 
cluster such as NGC 6402, which appears to be missing a normally well-populated
stellar group, may also provide insight into the connection between light 
element composition and HB location.

\begin{figure}
\includegraphics[width=\columnwidth]{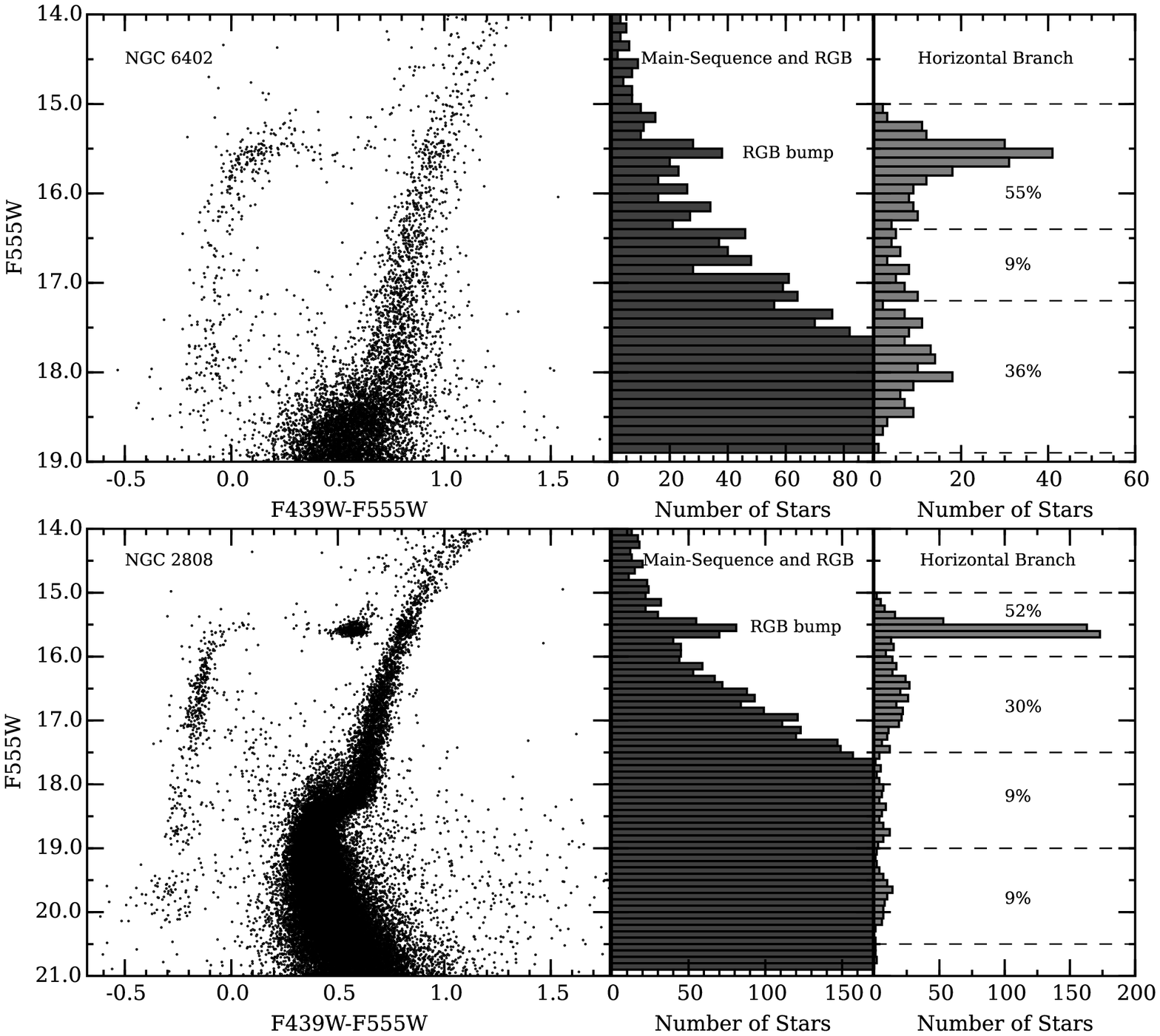}
\caption{F555W versus F439W-F555W CMDs are shown for NGC 6402 (top) and
NGC 2808 (bottom) using data from \citet{Piotto02}.  Both clusters have similar
metallicity and age values but exhibit very different horizontal branch
morphologies.  The luminosity functions to the right of each CMD are for the
main-sequence and RGB populations (middle panels; dark grey) and horizontal
branch (right panels; light grey) sequences.}
\label{fig:f13}
\end{figure}

An RGB star's detailed composition is closely tied to its eventual 
location on the HB \citep[e.g.,][]{Marino11_m4,Gratton11,Gratton13,Gratton14,
Marino14,Gratton15}.  The coolest HB stars frequently exhibit abundance 
patterns similar to those of the P1 RGB group while warmer HB stars tend to 
have lower O and Mg abundances along with higher N, Na, and Al abundances.  
However, the exact distribution is not the same for all clusters.  For example,
red HB stars show clear signatures of light element abundance variations in 
some clusters \citep[e.g.,][]{Gratton13,Marino14} but exhibit homogeneous 
compositions in others \citep[e.g.,][]{Marino11_m4,Gratton13}.  In some cases,
red HB, RR Lyrae, and/or the coolest blue HB stars may all contain P1 
composition stars \citep[e.g.,][]{Marino11_m4,Gratton15}.  Interestingly, the 
only stable observation is that for clusters with extended blue HBs the most 
polluted HB stars are found at the highest temperatures.  The generally 
accepted explanation is that He enhancement, in addition to cluster age and 
metallicity, drive the high temperature tail of a cluster's blue HB morphology 
\citep[e.g.,][]{Dantona02,Gratton10,Milone14}.

With these observations in mind, Fig.~\ref{fig:f13} compares the RGB and HB 
morphologies of NGC 2808 and NGC 6402, which as noted previously have similar 
ages, masses, metallicities\footnote{In addition to direct spectroscopic 
measurements, the similar red HB and RGB bump luminosities shown in 
Fig.~\ref{fig:f13} further support both clusters exhibiting comparable mean 
metallicities.}, and to some extent light element compositions.  However, we 
note that despite these similarities the detailed HB morphologies are 
different.

In Fig.~\ref{fig:f13}, the most obvious difference between the two clusters is 
the paucity of red HB stars in NGC 6402.  The red HB fraction for NGC 6402 is 
$\la$ 10 per cent \citep{ContrerasPena13} but exceeds 50 per cent for NGC 2808.
Although some P1 stars may reside on the red HB in NGC 6402, the 10 per cent 
fraction is too low to account for the 34 per cent P1 ratio found here.  
Therefore, a large fraction of the P1 and P2 stars in NGC 6402 must be shifted 
to the RR Lyrae region and cool end of the blue HB sequence.  In 
Fig.~\ref{fig:f13} this corresponds to the turn over region near F439W$-$F555W 
$\sim$ 0.5 mag., which leads us to conclude that the P1 and P2 stars in 
NGC 6402 reach the the zero age HB with lower envelope masses than their 
NGC 2808 counterparts.  The exact reason for this discrepancy is not clear but 
could result if NGC 6402 stars lose mass on the RGB more efficiently, have 
higher \emph{ab initio} He abundances (implying a shorter evolution time), or 
if NGC 6402 is considerably older.

Fig.~\ref{fig:f13} also highlights that the nearly vertical portions along the 
extended blue HBs also differ between the two clusters.  For NGC 6402, 36 per 
cent of the HB stars belong to the extreme blue tail with only 9 per cent 
residing at intermediate luminosities.  However, NGC 2808 has about 30 per cent
of its stars at intermediate luminosities and only $\sim$20 per cent on the 
extreme blue tail.  For NGC 2808, the ratio of intermediate to extreme (plus 
blue hook) HB stars is roughly comparable to the I1$+$I2 and E population 
fractions given in \citet{Carretta15}, respectively. 

\begin{figure*}
\includegraphics[width=\textwidth]{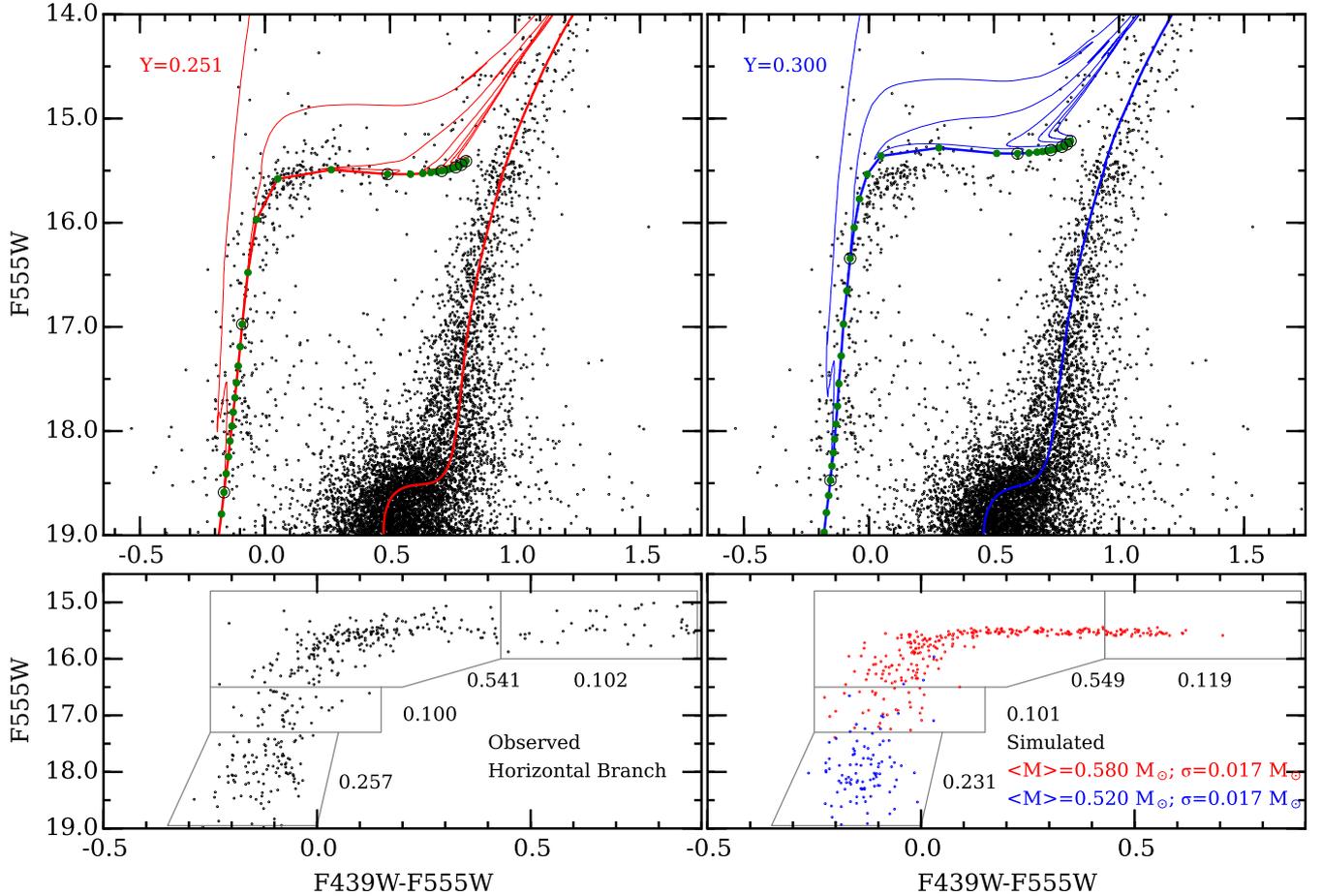}
\caption{\emph{Top:} F555W versus F439W-F555W CMDs are shown for NGC 6402 with
11.5 Gyr, $\alpha$-enhanced, [Fe/H] = $-$1 BaSTI \citep{Pietrinferni04}
isochrones and horizontal branch models shown for He abundances of Y = 0.251
(left panel; red) and Y = 0.300 (right panel; blue).  The solid green circles
on the zero-age horizontal branch represent masses ranging from about
0.8 M$_{\sun}$ to 0.48 M$_{\sun}$.  The points with larger open black circles
span, from blue to red, 0.5 to 0.8 M$_{\sun}$ in 0.05 M$_{\sun}$ increments.
These points also have full evolutionary sequences drawn as light red and
blue lines.  \emph{Bottom:} The left panel shows only the horizontal branch
stars with the numerical values indicating fractional populations within each
outlined region.  The right panel shows a simulated horizontal branch assuming
mean horizontal branch masses of 0.58 M$_{\sun}$ ($\sigma$ = 0.017 M$_{\sun}$)
and 0.52 M$_{\sun}$ ($\sigma$ = 0.017 M$_{\sun}$) for the He-normal (first
generation) and He-rich (second generation) stars, respectively.  The
simulations used the P1$+$P2 and E population ratios to determine the number
of objects in the He-normal and He-rich groups.}
\label{fig:f14}
\end{figure*}

The combined observations indicate that at [Fe/H] $\sim$ $-$1.15 the I1/I2 
composition globular cluster stars scatter along the upper/middle portion of 
the blue HB in CMDs using the F439W and F555W filters while those of the E 
population reach the highest temperatures.  The paucity of I1/I2 stars in 
NGC 6402 therefore seems a likely explanation for the small fraction of HB 
stars with 16 $\la$ F555W $\la$ 17 mag.  Furthermore, the comparatively high
fraction of extreme blue HB stars in NGC 6402 suggests that the cluster was 
able to form an unusually large fraction of He-enhanced stars with minimal
dilution.  

Fig.~\ref{fig:f14} shows that the full HB morphology of NGC 6402 may be modeled
relatively well using just two populations with different mean masses and He 
abundances but similar mass dispersions.  We selected Y = 0.3 for the 
He-enhanced population since the zero-age HB locus fit the level of the 
brightest HB stars.  However, we note that NGC 2808 may have $\Delta$Y values 
as high 0.13 \citep[e.g.,][]{Milone15_n2808}, and the combined HB and 
composition data suggest that NGC 6402 likely has $\Delta$Y $\sim$ 0.1.  This 
level of He enrichment could also fit the NGC 6402 HB as long as the most 
He-rich stars reside on the vertical portion of the blue HB. 

Regardless of the exact $\Delta$Y value, the HB evolutionary tracks in 
Fig.~\ref{fig:f14} show that the most extreme blue HB stars will not evolve to 
ascend the AGB.  As a result, we expect that like several other clusters, 
including NGC 2808 \citep{Wang16}, the AGB of NGC 6402 will be missing the most
Na-rich stars.  Differential reddening and field contamination make number 
counts of RGB and AGB stars in NGC 6402 difficult, but the cleaned CMD in 
Fig. 5 of \citet{ContrerasPena18} suggests the AGB is sparsely populated. 

\subsubsection{NGC 2419 and NGC 6402}

A comparison of the NGC 6402 CMD shown in Fig.~\ref{fig:f13} with the NGC 2419
CMD presented in Fig.~2 of \citet{DiCriscienzo11} reveals a striking similarity
between the two clusters.  The HBs of both clusters are numerically dominated 
by luminous blue HB stars and extreme blue HB stars, but each possesses a 
significant paucity of intermediate luminosity blue HB stars.  In fact, the 
distributions of luminous blue HB, intermediate, and extreme HB stars are 
nearly identical with NGC 6402 having 55, 9, and 36 per cent ratios and 
NGC 2419 exhibiting 50, 9, and 41 per cent fractions, respectively 
\citep{DiCriscienzo15}.

The similar HB distributions of NGC 6402 and NGC 2419 are particularly 
surprising given that the two clusters are separated in metallicity by about a
factor of 10.  Similar to the results presented here for NGC 6402, 
\citet{Cohen12} and \citet{Mucciarelli15} have also shown that NGC 2419 
exhibits a strongly bimodal light element distribution with dominant 
populations of P1$+$P2 and E stars but a paucity of intermediate composition 
stars.  Given suspicions that both NGC 6402 \citep[e.g.,][]{ContrerasPena18} 
and NGC 2419 \citep[e.g.,][]{Cohen12} may be the nuclear cores of former dwarf
galaxies, it is possible that these clusters trace similar modes of star 
formation but at different metallicities.  If true then similar clusters, both
within and outside the Galaxy, may be identified by the existence of extended
but strongly bimodal blue HB distributions.

\subsection{NGC 6402 Chemical Composition Implications for Cluster Formation}

As mentioned in Section 1, the detailed processes by which globular clusters
form and undergo self-enrichment are unclear.  Most models assume that 
so-called first generation stars are the first to form from the cluster's 
natal gas \citep[e.g.,][]{Dantona16}.  However, some models predict the 
opposite behavior and assume that the O/Mg-poor and N/Na/Al-rich stars form 
first \citep{Marcolini09}.  Establishing the order in which various globular 
cluster populations form would provide deep insight into the physical processes
and pollution sources that shape the composition patterns observed today.  In
this context, the peculiar light element pattern of NGC 6402 may shed light on 
the cluster formation process.

Several lines of evidence indicate that P1 stars are the first to form.  For 
example, Fig.~\ref{fig:f7} emphasizes that all old Galactic globular clusters 
examined to date have substantial populations of P1 stars.  Fig.~\ref{fig:f7}
also shows that lower mass clusters, which tend to exhibit smaller composition
variations and therefore less self-enrichment, have the largest P1 fractions.
We also note that clusters composed solely of intermediate or extreme 
composition stars have never been found and that Rup 106, the only confirmed 
single population cluster \citep{Villanova13,Dotter18}, possesses solely low-Na 
stars.  Additionally, a small number of N-rich stars, which may have originated
in globular clusters, have been found in the Galactic bulge and halo 
\citep[e.g.,][]{Martell16,Schiavon17}, but the ratios are not high enough to 
suggest that clusters composed solely of ``second generation" stars ever 
existed.

A more significant controversy surrounds the formation of the
intermediate and extreme composition stars.  Since most proposed pollution 
sources do not inherently produce the observed light element
abundance patterns, models generally require some level of dilution between 
processed and pristine gas \citep[e.g.,][]{Dercole11,Dantona16}.  In fact,
many models attempt to fit the light element distributions shown in 
Fig.~\ref{fig:f8} by mixing only P1 and E composition gas or P1 and model 
ejecta gas.  Although these models produce reasonable fits for some clusters, 
recent analyses have shown that simple dilution models are insufficient to 
simultaneously fit all of the populations in at least NGC 6752, NGC 2808, and 
NGC 5986 \citep{Carretta12_6752,Carretta18,Johnson17_5986}.  The inability of 
single dilution models to fit these clusters suggests that more than one 
pollution source was in operation (see also Section 5.4).  In other words, 
intermediate and extreme globular cluster stars need not follow a common 
dilution function, form at the same time, nor originate from the same pollution
source.

The large gaps in NGC 6402's [O/Fe], [Na/Fe], and [Al/Fe] abundances support 
the notion that most globular clusters are enriched by distinct, but separate,
events rather than a simple dilution of single composition pristine and 
polluted gas.  Since the I1/I2 stars are missing, or at least a minority 
population, in NGC 6402 but represent a large fraction of most clusters, the 
present data suggest that I1/I2 stars are the \emph{last} group to form.  
Delayed formation of intermediate composition stars has been proposed in the 
past \citep[e.g.,][]{Valcarce11,Dantona16,Bekki17_discrete,Kim18}, but NGC 6402 
represents the first direct evidence supporting such a hypothesis.  The 
missing I1/I2 group further suggests that NGC 6402 may be a special case where 
star formation terminated prematurely.

If NGC 6402 is a preserved snapshot of globular cluster formation then the 
relationships between the P1, P2, and E populations are particularly 
intriguing.  Fig.~\ref{fig:f7} shows that NGC 6402 has a large, but not 
unusual, number of P1 stars, and the available evidence is in-line with the 
previously stated assumption that these were among the first stars to form.  In
fact, the I1/I2 gap may completely rule out all formation scenarios where P1 
stars form through any type of dilution event.  

\begin{figure*}
\includegraphics[width=\textwidth]{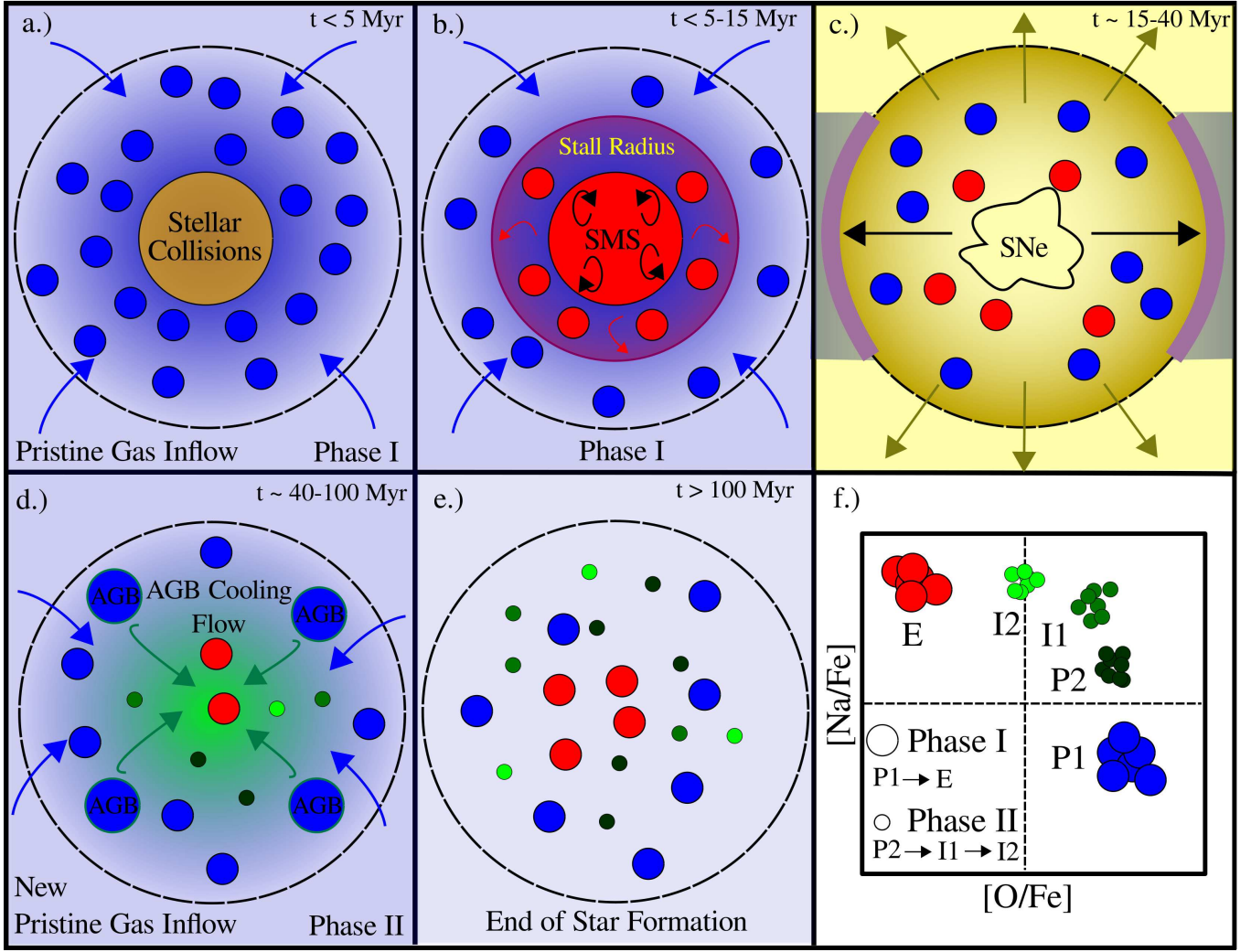}
\caption{An illustration of the proposed globular cluster formation model.
Note that the schematic is not to scale.  Panel (a) illustrates the
formation of P1 first generation stars (blue filled circles) from the initial
cold gas present in the cluster.  P1 stars are assumed to form at all radii and
follow a typical cluster density profile.  For sufficiently massive and dense
cluster cores, stellar collisions are likely to occur and result in the
production of a centrally concentrated supermassive star (SMS;
$\ga$10$^{\rm 3}$ M$_{\sun}$).  Following \citet{Gieles18} and
\citet{Denissenkov14}, panel (b) indicates that the fully convective SMS will
eject gas that has been processed at high temperatures and matches the
composition of E stars (red filled circles).  \citet{Gieles18} predict that the
SMS gas will be retained within a central ``stall radius" (where the SMS wind
pressure matches the intracluster gas pressure) and lead to the formation of E
composition stars.  After $\sim$10$-$15 Myr, panel(c) illustrates that the
first Type II SNe will explode and drive out the intracluster gas, which may
also coincide with the dissolution of the SMS.  Following the \citet{Dercole16}
bubble model, a cluster forming within a dense disk may eject most of its gas
out of the plane but retain some local gas in a shell within the plane
(thick purple lines).  Panel (d) indicates that after the SN era is over the
cluster will accumulate more pristine gas and also develop a central cooling 
flow of intermediate mass AGB ejecta.  Mixing of pristine and AGB gas will 
produce the P2, I1, and I2 intermediate composition stars (small filled green 
circles).  The formation order will depend on the local ratio of 
pristine-to-polluted gas, but the results from NGC 6402 indicate that the P2
stars can form first.  Panel (e) shows the cluster composition after
star formation has ceased, and panel (f) is a schematic illustration of the
[O/Fe] and [Na/Fe] abundances for each population.}
\label{fig:f15}
\end{figure*}

In Section 5.4, we used the different composition patterns of NGC 6402 and 
NGC 2808 to illustrate a fundamental disconnect between the formation of E and 
P2 population stars.  Specifically, we concluded that the light element 
patterns of the two clusters could not be explained unless the P2 and E groups 
formed from gas that originated from different sources.  The data therefore 
support a model in which E composition stars form as an independent group and 
where the P2, I1, and I2 stars are the only populations requiring significant
dilution.  

If AGB stars are important contributors of polluted gas, then the 
\citet{Dercole08} model indicates that chemical enrichment via AGB winds could
produce a cooling flow that funnels N/Na/Al-rich and O/Mg-poor gas toward the
cluster core and result in the formation of intermediate composition stars,
after mixing with pristine gas (see Fig.~\ref{fig:f15}).  The NGC 6402 data 
suggest that the P2 stars form before the I1 and I2 groups, which would invert 
the middle formation order presented by \citet{Dantona16}.  If P2 stars form 
before I1/I2 stars in all clusters then the data would indicate that a large 
reservoir of pristine gas must be present before the intermediate composition 
stars can start forming, which presumably occurs after the initial Type II 
supernova (SN) era.  In this scenario, the I1/I2 composition stars would then 
form from gas experiencing a decreasing ratio of pristine-to-polluted material 
with time.  However, the P2$\rightarrow$I1$\rightarrow$I2 sequence need not be 
universal since the I1/I2 stars could form before the P2 group if the polluting
gas reaches a high enough density to initiate star formation before a 
substantial amount of pristine gas is re-accumulated after the SN era.

A delay of several tens of Myrs between the production of P1 and P2 stars seems
required to provide enough time for the latter pollution source(s) (AGB stars?)
to evolve.  The ``bubble" model described in \citet{Dercole16} predicts a delay
of $>$ 40 Myr between the gas expulsion phase initiated by Type II supernovae 
(SNe) and the re-collapse/accretion of pristine material.  Fig.~\ref{fig:f12}
showed that AGB stars may be able to produce the P2, I1, and I2 composition
stars, but the current models do not easily explain the E populations that are
present in only the most massive clusters.  Therefore, we hypothesize that the 
E stars require special circumstances for formation (see Fig.~\ref{fig:f15}),
and posit that both the P1 and E populations are formed very early in a 
cluster's lifetime, perhaps within the first 5$-$10 Myr.  We follow the hybrid 
pollution scenario presented in \citet{Sills10} and suggest that supermassive 
star enrichment may provide a plausible explanation for the E population.

The formation of supermassive stars may be prominent in the dense, gas-rich 
environments present in some early globular clusters, and \citet{Gieles18} 
showed that the required conditions for these objects to form may be limited
to the most massive clusters ($\ga$ 10$^{\rm 6}$ stars) with high gas 
accretion rates ($\ga$ 10$^{\rm 5}$ M$_{\sun}$ Myr$^{\rm -1}$).  With masses
exceeding 10$^{\rm 3}$ M$_{\sun}$, supermassive stars would be able to provide
processed material to a cluster interstellar medium within the first few Myr,
and \citet{Denissenkov14} showed that such stars could produce gas matching the
composition of E stars.  The high He and N abundances of the E composition gas
would permit efficient cooling \citep{Herwig12} and further favor rapid and 
early star formation.  Additional He and N enhancements could also be provided
from the winds of massive WN type Wolf-Rayet stars and Luminous Blue Variables
as well \citep[e.g.,][]{Dufour97,Smith04,Smith06}.  \citet{Gieles18} also noted
that winds from a central supermassive star may be confined within a small 
``stall radius", where the wind pressure and ambient cluster pressure are 
equal, which could maintain an environment that largely excluded mixing with 
pristine gas and promoted the formation of nearly pure E composition stars 
(see Fig.~\ref{fig:f15}).  Since lower mass clusters would not form 
supermassive stars, the production of E group stars would be limited to the 
most massive clusters with the highest gas accretion rates.  Furthermore, 
confinement of the supermassive star winds to small cluster radii would provide
a natural explanation for the frequently observed central concentration of E 
composition stars \citep[e.g.,][]{Carretta10_3201,Johnson10,Lardo11,Johnson12}.

Regardless of the specific mechanism that produces E composition stars, 
kinematic studies support the notion that the E populations represent distinct
formation events.  In addition to being preferentially located near the cores 
of several clusters, the E stars can also exhibit faster rotation 
\citep{Cordero17} and possibly smaller line-of-sight velocity dispersions 
\citep[e.g.,][]{Bellazzini12,Kucinskas14,Cordero17} than other cluster 
sub-populations.  In this light NGC 6402 is no exception since the velocity 
dispersions for the P1, P2, P1$+$P2, and E populations are 11.4, 8.0, 9.5, and 
4.6 km s$^{\rm -1}$, respectively.  However, the number of targets and radial 
sampling in the present data set are too small to draw any firm conclusions 
regarding kinematic and/or radial distribution signatures that may be unique to
the E population in NGC 6402.

The scenario outlined in Fig.~\ref{fig:f15} indicates that the E and I2 groups 
are both composed of nearly pure polluted material except that each would 
originate from a different source.  Although direct comparisons between E and 
I2 stars are sparse in the literature, we note that Li observations may provide
some support of different origins for the E and I2 groups.  If intermediate and 
extreme composition stars formed from gas burned at high temperatures then 
lithium's low destruction temperature ($\sim$2.5$\times$10$^{\rm 6}$ K) means 
A(Li) should strongly decrease in intermediate and extreme stars.  Li should 
also be anti-correlated with elements such as Na and Al, but \citet{Dorazi15} 
found that Li abundances are generally invariant as a function of [Na,Al/Fe]. 
On the other hand, \citet{Lind09}, \citet{Dorazi15}, and \citet{Mucciarelli18} 
have shown that E composition stars in particular are Li-depleted, which 
follows the hypothesis that the P2, I1, and I2 groups in globular clusters are 
polluted by different sources (where some Li may have been produced) than the E
stars.  We note that some authors do find Li-Na anti-correlations that are 
not restricted to just the E populations \citep[e.g.,][]{Pasquini05,
Bonifacio07} so it is possible that the production (or not) of Li is cluster
specific and depends on the particular polluting source(s) mixing with pristine
gas.

The model shown in Fig.~\ref{fig:f15} attempts to alleviate many of the 
cluster formation problems outlined in Fig.~6 of \citet{Bastian18}, 
particularly with regard to detailed abundance patterns, the discrete nature
of various sub-populations, and relationships between composition variations 
and cluster mass.  On the other hand, the mass budget problem may persist
and the present model still requires a full enrichment timescale of at least 
40$-$100 Myr, which seems to contrast with observations of many nearby young 
massive clusters that appear to become gas free after only a few Myr 
\citep[e.g.,][]{Bastian14,Hollyhead15}.  The lack of an Fe spread in clusters 
that continued to form stars after the SN era could also be a concern.  
However, we note that the products of SN explosions are not guaranteed to 
contaminate the intracluster gas \citep[e.g.,][]{Tenorio15} nor completely 
clear a cluster of natal gas \citep[e.g.,][]{Dercole16}.

It is possible that many of the investigated young massive clusters are not 
actually close analogs of early globular clusters, and/or that the early 
Universe environment favored higher gas densities than are common today 
\citep[e.g.,][]{Elmegreen17,Pfeffer18}.  Interestingly, the ``supernebula" 
region located in NGC 5253 may provide an alternative analog to early globular 
cluster formation as it contains at least two compact, young ($\sim$1$-$3 Myr), 
deeply embedded, and massive ($\sim$2$\times$10$^{\rm 6}$ M$_{\sun}$ combined)
star clusters with star formation efficiencies of $\sim$30$-$70 per cent 
\citep[e.g.,][]{Beck96,Calzetti15,Turner15,Turner17,Miura18}.  This region also
contains several thousand O stars, many Wolf-Rayet stars, perhaps a dozen or
more very massive stars exceeding 150 M$_{\sun}$, and may also have a top-heavy 
initial mass function \citep[e.g.,][]{Turner15,Turner17}.  The extremely high
densities experienced by the supernebula may be driven in part by the infall
of a diffuse H{\sevensize I} cloud \citep{Lopez12,Westmoquette13}, and it is 
possible that this environment somewhat resembles early globular cluster
formation occurring during the dominant galaxy assembly and star formation
epoch at high redshift.

Although the supernebula clusters are still embedded in their natal clouds, 
several analyses have detected N-enhanced gas, along with an anti-correlation
between log(O/H) and log(N/O), associated with the supernebula region that 
could be connected to pollution from Wolf-Rayet stars, Luminous Blue Variable 
stars, and/or very massive ($>$ 100 M$_{\sun}$) stars \citep{Walsh87,
Kobulnicky97,Lopez07,Monreal10,Westmoquette13,Smith16}.  He enhancements have 
also been detected \citep{Lopez07} but are so far unconfirmed 
\citep{Monreal13}.  \citet{Cohen18} showed that strong radiative cooling 
within the NGC 5253 supernebula region is likely stalling stellar winds and 
preventing the formation of significant cluster-scale outflow, which could 
enable prolonged star formation from chemically enriched gas.  Even if the 
pollution sources are not exactly the same as those in early Universe massive 
clusters, the results from NGC 5253 indicate that enriched gas from massive 
stars can be produced quickly and retained for at least a few Myr within an 
extremely dense proto-cluster region\footnote{We note that if the winds from
Wolf-Rayet stars, Luminous Blue Variables, and other massive stars are enriched
almost exclusively in He/N then these sources may explain the suspected
He/N-enhanced first generation stars noted by \citet{Milone18_1gspread} and 
\citet{Lardo18}.  Following \citet{Silich17}, the retention (or not) of early 
ejecta from high mass stars may be strongly tied to the mass and density of the
proto-cluster environment, which is at least qualitatively in agreement with 
the observed correlation between present-day cluster mass and the estimated
He-spread among first generation stars shown in 
\citet[][their Figure 19]{Milone18_1gspread}.  The strong density requirement
for retaining stellar winds shown by \citet[][their Figure 1]{Silich17}, 
especially at high star formation efficiency, may explain why only some 
clusters with similar masses exhibit He/N spreads among their first generation
populations \citep{Milone18_1gspread,Lardo18}.}.  Furthermore, \citet{Silich17}
showed that massive compact clusters, such as those within the NGC 5253 
supernebula, will likely retain some natal gas through the SN era.  This mode
of gas retention supports the possibility of a delayed star formation 
epoch featuring pollution from longer lived sources, such as AGB stars 
\citep[see also][]{Kim18}.

To summarize (see also Fig.~\ref{fig:f15}), we hypothesize that monometallic 
globular clusters form in two distinct stages.  During phase one, which likely 
occurs during the first few Myr, all globular clusters form P1 composition 
stars at a variety of radii.  In the most massive systems with the highest gas
accretion rates, central supermassive stars form via runaway collisions 
\citep[e.g.,][]{Sills10,Gieles18} and pollute cluster cores with the products 
of high temperature proton-capture burning\footnote{The proposed cluster 
formation model does not necessarily require a central $>$ 10$^{\rm 3}$ 
M$_{\sun}$ star, especially since such supermassive stars have never been 
observed.  Instead, the only requirements are that a mechanism exists to 
produce very O/Mg-poor and He/Na/Al-rich gas on a time scale $\la$ 10 Myr, that
this process favors dense, high cluster mass environments, and that most of the
material is confined to the cluster core.}.  Strong radiative cooling and the 
stalling of stellar winds promote rapid star formation, significant gas 
retention, and inhibit the mixing of polluted and pristine gas, which produces
a centrally concentrated population of E composition stars.  Following the 
model presented by \citet{Dercole16}, core-collapse SNe halt star formation for
$\sim$15$-$40 Myr until enough pristine gas can be accumulated to start the 
second, delayed formation phase.  After $\sim$40 Myr, the \citet{Dantona16} 
model indicates that $<$ 8 M$_{\sun}$ AGB stars can start contributing 
processed gas and mixing with pristine material.  The I1/I2 gap in NGC 6402 
suggests that stars with the highest pristine-to-polluted gas ratios form first
(P2 group), and that the I1 and I2 populations form later as the pristine gas 
fraction drops.  Presumably, star formation is permanently halted either by the
mechanical ejection of all gas (e.g., Type Ia SNe) or when the remaining gas 
density falls below a critical limit.  

\section{Summary}

This paper presents radial velocity and/or chemical abundance measurements for
41 RGB stars in the massive Galactic globular cluster NGC 6402 using data 
obtained with the M2FS instrument in high resolution mode.  We find a mean
cluster heliocentric radial velocity of $-$61.1 km s$^{\rm -1}$ with a 
dispersion of 8.5 km s$^{\rm -1}$.  Additionally, we measured 
$\langle$[Fe/H]$\rangle$ = $-$1.13 dex and a small star-to-star dispersion of
0.05 dex.

The overall composition properties of NGC 6402 are similar to those of other
inner Galaxy clusters such that NGC 6402 is generally $\alpha$-enhanced,
has approximately solar [Cr,Ni/Fe] ratios, and exhibits a moderately low
$\langle$[La/Eu]$\rangle$ = $-$0.08 dex.  None of the elements heavier than
Ca show any evidence supporting significant abundance spreads.

The light elements O, Na, Mg, Al, and Si exhibit clear (anti-)correlations that
follow the typical patterns observed in other massive Galactic clusters.  
We also found some evidence supporting a correlation between [Al/Fe] and 
[Ca/Fe], but a comparison with NGC 2808 suggests that the Al-Ca correlation may
be spurious.  Interestingly, the [O/Fe], [Na/Fe], and [Al/Fe] data exhibit 
large abundance gaps that indicate a missing or minority population of 
intermediate composition stars.  

Statistical analyses and an examination of NGC 6402's HB morphology suggest 
that intermediate composition stars likely represent $<$ 10 per cent of the 
total cluster population.  However, the extreme composition stars constitute 
$>$ 25 per cent of our sample, and their likely location on the high 
temperature tail of the blue HB may correlate with a high He abundance as well.

A detailed light element composition comparison between NGC 6402 and NGC 2808 
indicates that the various sub-populations in NGC 6402 formed from gas that was
burned at different temperatures.  This result suggests that in NGC 6402, and 
probably most clusters, the full extension of light element abundance 
variations is likely driven by pollution from multiple sources.  In agreement 
with some theoretical models, the observed [O/Fe], [Na/Fe], and [Al/Fe] gaps in
NGC 6402 seem to indicate that intermediate composition stars are the last 
group to form.  As a result, we followed previous work to develop a qualitative
formation model in which monometallic globular clusters experience two distinct
phases of star formation.

In the first rapid formation stage, a significant fraction of pristine gas is 
converted into stars that exhibit ``halo-like" abundance patterns.  During this
first formation phase, high mass clusters such as NGC 6402 may also be able to 
drive the production of supermassive stars near their cores, and as a result
produce centrally concentrated populations of extreme composition stars.  
Recent examinations of young massive clusters indicate that they may become 
gas free after only $\sim$5$-$10 Myr, and we posit that the most extreme 
composition stars in globular clusters probably form within this approximate
time frame.  Rapid star formation may be further aided by the enhanced He and N
abundances of E composition gas, which has been shown to permit efficient 
cooling.  An examination of the supernebula region in NGC 5253 suggests that 
rapid chemical enrichment is possible within very dense proto-cluster cores, 
and wind stalling may further promote prolonged star formation from chemically
enriched gas.  Following the notation of \citet{Carretta15}, clusters like 
NGC 6402 would contain mostly P1 and E stars at the end of the first formation 
stage.  In contrast, lower mass clusters would not be able to form central 
supermassive stars and as a result would only have P1 composition stars at the 
end of the first formation phase.  In both cases, star formation would likely 
cease for $\sim$10$-$40 Myr while core-collapse SNe explode and drive out gas.

After $\sim$40 Myr, a second star formation event is initialized when the 
re-accumulated gas density reaches some critical value.  Although present-day 
young massive clusters do not seem to retain natal gas 
after 40 Myr, \citet{Silich17} showed that gas retention is possible for very 
compact proto-clusters.  A combination of polluted gas, perhaps driven by 
various cluster sources such as intermediate mass AGB stars, and pristine gas 
would then mix to produce the variety of intermediate composition 
stars found in nearly all globular clusters.  The time scale of gas 
accumulation and the rate at which the pristine-to-polluted gas ratio changes 
would then govern the exact distribution of intermediate composition stars, 
which would likely vary from cluster-to-cluster.  For NGC 6402, the missing 
I1/I2 populations suggest that P2 stars can form first during this 
later phase.  Since P2 stars have high pristine-to-polluted gas ratios, the 
NGC 6402 data indicate that the delayed star formation phase is dependent on 
the re-accumulation of pristine gas for initialization.  In this case, the 
process of forming P2$\rightarrow$I1$\rightarrow$I2 stars would make sense if 
the cluster's overall ratio of pristine-to-polluted gas decreased with time.  
For all clusters, star formation would cease when the gas density became too 
low or was otherwise removed (e.g., Type Ia SNe).  In the case of NGC 6402, the
small or missing I1 and I2 groups suggest that star formation was 
prematurely halted and that the cluster may represent a frozen snap shot of 
the chemical enrichment process.

%
%
%


\section*{Acknowledgements}

This research has made use of NASA's Astrophysics Data System Bibliographic
Services.  This publication has made use of data products from the Two
Micron All Sky Survey, which is a joint project of the University of
Massachusetts and the Infrared Processing and Analysis Center/California
Institute of Technology, funded by the National Aeronautics and Space
Administration and the National Science Foundation.  We thank Jean Turner for
helpful discussions related to NGC 5253.  C.I.J. gratefully acknowledges 
support from the Clay Fellowship, administered by the Smithsonian Astrophysical
Observatory.




\bibliographystyle{mnras}
\bibliography{references}



\begin{landscape}
\begin{table}
  \caption{Table 1: star identifiers, sky coordinates, photometry, radial 
  velocities, and model atmosphere parameters are provided for all member and 
  non-member stars observed for this program.}
  \label{tab:basic_params}
  \begin{tabular}{cccccccccccc}
  \hline
  Star Name	&	RA	&	DEC	&	J	&	H	&	K$_{\rm S}$	&	RV$_{\rm Helio.}$	&	RV Error	&	T$_{\rm eff}$	&	log(g)	&	[Fe/H]	&	$\xi$$_{\rm mic.}$\\
  (2MASS)	&	(J2000)	&	(J2000)	&	(mag.)	&	(mag.)	&	(mag.)	&	(km s$^{\rm -1}$)	&	(km s$^{\rm -1}$)	&	(K)	&	(cgs)	&	(dex)	&	(km s$^{\rm -1}$)\\
  \hline
  \multicolumn{12}{c}{Cluster Members}\\
  \hline
2M17372576$-$0314077	&	264.357353	&	$-$3.235491	&	11.368	&	10.505	&	10.246	&	$-$64.3	&	0.5	&	4175	&	1.00	&	$-$1.13	&	1.90	\\
2M17372832$-$0314210	&	264.368037	&	$-$3.239184	&	11.732	&	10.850	&	10.643	&	$-$63.4	&	0.7	&	...	&	...	&	...	&	...	\\
2M17373006$-$0314502	&	264.375283	&	$-$3.247291	&	11.640	&	10.848	&	10.545	&	$-$50.2	&	0.4	&	...	&	...	&	...	&	...	\\
2M17373048$-$0315530	&	264.377033	&	$-$3.264747	&	10.967	&	10.106	&	9.856	&	$-$60.5	&	2.0	&	...	&	...	&	...	&	...	\\
2M17373178$-$0313592	&	264.382423	&	$-$3.233112	&	12.188	&	11.382	&	11.167	&	$-$58.7	&	0.6	&	4425	&	1.40	&	$-$1.14	&	1.65	\\
2M17373225$-$0311152	&	264.384408	&	$-$3.187576	&	10.966	&	10.071	&	9.817	&	$-$56.9	&	0.5	&	4075	&	0.85	&	$-$1.13	&	1.95	\\
2M17373243$-$0313213	&	264.385131	&	$-$3.222595	&	12.209	&	11.394	&	11.175	&	$-$64.1	&	0.9	&	4375	&	1.25	&	$-$1.14	&	1.65	\\
2M17373305$-$0311386	&	264.387709	&	$-$3.194075	&	11.951	&	11.185	&	10.947	&	$-$63.3	&	0.5	&	4375	&	1.20	&	$-$1.14	&	1.75	\\
2M17373325$-$0315382	&	264.388560	&	$-$3.260634	&	10.987	&	10.040	&	9.772	&	$-$63.7	&	0.4	&	4075	&	1.05	&	$-$1.15	&	1.95	\\
2M17373336$-$0315103	&	264.389006	&	$-$3.252870	&	12.057	&	11.234	&	10.965	&	$-$60.0	&	0.5	&	4300	&	1.00	&	$-$1.09	&	1.55	\\
2M17373351$-$0314396	&	264.389637	&	$-$3.244342	&	10.951	&	10.014	&	9.753	&	$-$66.0	&	0.6	&	4035	&	0.60	&	$-$1.17	&	1.85	\\
2M17373366$-$0312495	&	264.390260	&	$-$3.213775	&	12.227	&	11.401	&	11.210	&	$-$45.5	&	0.3	&	4425	&	1.40	&	$-$1.11	&	1.70	\\
2M17373382$-$0315285	&	264.390933	&	$-$3.257925	&	12.310	&	11.470	&	11.268	&	$-$61.7	&	0.7	&	4475	&	1.50	&	$-$1.17	&	1.70	\\
2M17373390$-$0313291	&	264.391283	&	$-$3.224769	&	10.983	&	10.043	&	9.774	&	$-$73.0	&	0.8	&	4065	&	0.45	&	$-$1.21	&	1.95	\\
2M17373395$-$0313138	&	264.391472	&	$-$3.220503	&	11.803	&	10.918	&	10.710	&	$-$68.1	&	0.6	&	4375	&	1.25	&	$-$1.10	&	1.85	\\
2M17373458$-$0315552	&	264.394097	&	$-$3.265336	&	12.087	&	11.258	&	11.036	&	$-$59.1	&	0.2	&	4500	&	1.60	&	$-$1.13	&	1.75	\\
2M17373460$-$0314057	&	264.394192	&	$-$3.234935	&	11.614	&	10.716	&	10.533	&	$-$43.7	&	0.4	&	4275	&	1.05	&	$-$1.18	&	1.60	\\
2M17373470$-$0315094	&	264.394601	&	$-$3.252625	&	11.579	&	10.750	&	10.503	&	$-$62.2	&	0.5	&	...	&	...	&	...	&	...	\\
2M17373606$-$0315251	&	264.400256	&	$-$3.256982	&	10.957	&	10.032	&	9.785	&	$-$50.3	&	0.2	&	4300	&	1.15	&	$-$1.22	&	1.75	\\
2M17373691$-$0313304	&	264.403825	&	$-$3.225115	&	11.637	&	10.796	&	10.529	&	$-$59.2	&	0.4	&	4315	&	1.10	&	$-$1.11	&	1.85	\\
2M17373796$-$0314120	&	264.408198	&	$-$3.236675	&	11.311	&	10.434	&	10.152	&	$-$73.7	&	0.8	&	4125	&	1.00	&	$-$1.21	&	1.80	\\
2M17373815$-$0314496	&	264.408965	&	$-$3.247136	&	11.726	&	10.899	&	10.629	&	$-$68.2	&	0.9	&	4350	&	1.35	&	$-$1.15	&	1.80	\\
2M17373864$-$0313445	&	264.411013	&	$-$3.229044	&	10.897	&	9.991	&	9.708	&	$-$64.7	&	0.4	&	4075	&	0.60	&	$-$1.25	&	1.80	\\
2M17373909$-$0313237	&	264.412881	&	$-$3.223258	&	12.128	&	11.286	&	11.079	&	$-$77.7	&	1.3	&	...	&	...	&	...	&	...	\\
2M17373921$-$0314264	&	264.413402	&	$-$3.240683	&	11.042	&	10.166	&	9.879	&	$-$54.0	&	0.1	&	4175	&	0.90	&	$-$1.18	&	1.85	\\
2M17373944$-$0314480	&	264.414374	&	$-$3.246685	&	12.308	&	11.478	&	11.274	&	$-$71.8	&	1.2	&	4475	&	1.55	&	$-$1.13	&	1.60	\\
2M17373961$-$0315269	&	264.415077	&	$-$3.257477	&	10.903	&	9.975	&	9.725	&	$-$53.0	&	0.2	&	4200	&	1.05	&	$-$1.17	&	2.05	\\
2M17374031$-$0315544	&	264.417992	&	$-$3.265132	&	10.895	&	9.983	&	9.713	&	$-$68.6	&	0.8	&	4150	&	0.95	&	$-$1.17	&	1.95	\\
2M17374053$-$0314592	&	264.418885	&	$-$3.249801	&	12.039	&	11.247	&	11.011	&	$-$62.8	&	0.7	&	4400	&	1.15	&	$-$1.17	&	1.60	\\
2M17374093$-$0315327	&	264.420575	&	$-$3.259104	&	10.904	&	9.991	&	9.728	&	$-$46.0	&	0.1	&	4150	&	0.65	&	$-$1.01	&	1.75	\\
2M17374103$-$0313167	&	264.420999	&	$-$3.221331	&	11.852	&	11.038	&	10.782	&	$-$59.4	&	0.2	&	4475	&	1.65	&	$-$1.14	&	1.85	\\
2M17374139$-$0313528	&	264.422492	&	$-$3.231334	&	12.354	&	11.530	&	11.274	&	$-$65.7	&	0.8	&	...	&	...	&	...	&	...	\\
2M17374163$-$0314213	&	264.423482	&	$-$3.239265	&	11.337	&	10.385	&	10.148	&	$-$58.1	&	0.4	&	4075	&	0.75	&	$-$1.06	&	1.75	\\
2M17374212$-$0315069	&	264.425504	&	$-$3.251931	&	11.846	&	11.110	&	10.743	&	$-$70.6	&	1.2	&	4375	&	1.45	&	$-$1.11	&	1.70	\\
2M17374273$-$0315238	&	264.428065	&	$-$3.256630	&	11.403	&	10.504	&	10.288	&	$-$60.7	&	0.3	&	4300	&	1.25	&	$-$1.09	&	1.90	\\
2M17374309$-$0314044	&	264.429546	&	$-$3.234561	&	10.920	&	10.006	&	9.778	&	$-$44.9	&	0.2	&	4150	&	1.00	&	$-$1.16	&	1.90	\\
2M17374421$-$0313013	&	264.434233	&	$-$3.217049	&	11.873	&	10.993	&	10.771	&	$-$55.4	&	0.2	&	4350	&	1.35	&	$-$1.01	&	1.75	\\
2M17374450$-$0314255	&	264.435418	&	$-$3.240421	&	11.985	&	11.133	&	10.906	&	$-$57.1	&	0.3	&	4350	&	1.35	&	$-$1.12	&	1.85	\\
2M17374518$-$0314040	&	264.438288	&	$-$3.234462	&	11.854	&	10.997	&	10.780	&	$-$66.7	&	0.8	&	4275	&	1.10	&	$-$1.13	&	1.70	\\
2M17374581$-$0314407	&	264.440914	&	$-$3.244651	&	11.441	&	10.578	&	10.318	&	$-$77.8	&	1.0	&	4250	&	1.35	&	$-$1.02	&	1.80	\\
2M17374610$-$0315079	&	264.442089	&	$-$3.252207	&	11.378	&	10.480	&	10.246	&	$-$55.1	&	0.5	&	4250	&	1.10	&	$-$1.13	&	1.85	\\
  \hline
  \multicolumn{12}{c}{Non-Members}\\
  \hline
2M17372950$-$0314302	&	264.372934	&	$-$3.241728	&	11.834	&	11.041	&	10.782	&	$-$7.7	&	1.0	&	...	&	...	&	...	&	...\\
  \hline
  \end{tabular}
\end{table}
\end{landscape}

\begin{landscape}
\begin{table}
  \caption{Table 2: abundance ratios and population designations for all member
stars.}
  \label{tab:abundances}
  \begin{tabular}{cccccccccccccc}
  \hline
  Star	&	Population	&	[Fe{\sevensize I}/H]	&	[Fe{\sevensize II}/H]	&	[O{\sevensize I}/Fe]	&	[Na{\sevensize I}/Fe]	&	[Mg{\sevensize I}/Fe]	&	[Al{\sevensize I}/Fe]	&	[Si{\sevensize I}/Fe]	&	[Ca{\sevensize I}/Fe]	&	[Cr{\sevensize I}/Fe]	&	[Ni{\sevensize I}/Fe]	&	[La{\sevensize II}/Fe]	&	[Eu{\sevensize II}/Fe]\\
  (2MASS)	&	&	(dex)	&	(dex)   &	(dex)   &	(dex)   &	(dex)   &	(dex)   &	(dex)   &	(dex)   &	(dex)   &	(dex)   &	(dex)   &	(dex)\\
  \hline
2M17372576$-$0314077	&	P2	&	$-$1.12	&	$-$1.13	&	0.60	&	0.05	&	0.42	&	0.59	&	0.27	&	0.27	&	0.17	&	$-$0.07	&	0.35	&	0.21	\\
2M17372832$-$0314210	&	...	&	...	&	...	&	...	&	...	&	...	&	...	&	...	&	...	&	...	&	...	&	...	&	...	\\
2M17373006$-$0314502	&	...	&	...	&	...	&	...	&	...	&	...	&	...	&	...	&	...	&	...	&	...	&	...	&	...	\\
2M17373048$-$0315530	&	...	&	...	&	...	&	...	&	...	&	...	&	...	&	...	&	...	&	...	&	...	&	...	&	...	\\
2M17373178$-$0313592	&	P1	&	$-$1.12	&	$-$1.15	&	0.53	&	$-$0.13	&	0.39	&	0.34	&	0.31	&	0.27	&	$-$0.03	&	$-$0.07	&	0.28	&	0.23	\\
2M17373225$-$0311152	&	E	&	$-$1.13	&	$-$1.13	&	$-$0.22	&	0.56	&	0.30	&	1.27	&	0.63	&	0.31	&	0.04	&	$-$0.15	&	0.16	&	0.42	\\
2M17373243$-$0313213	&	P2	&	$-$1.13	&	$-$1.15	&	0.28	&	0.16	&	...	&	0.46	&	0.34	&	0.38	&	0.08	&	$-$0.19	&	0.11	&	0.39	\\
2M17373305$-$0311386	&	P2	&	$-$1.14	&	$-$1.14	&	...	&	0.31	&	0.41	&	0.57	&	0.35	&	0.41	&	0.02	&	$-$0.10	&	0.28	&	0.32	\\
2M17373325$-$0315382	&	P1	&	$-$1.14	&	$-$1.16	&	0.62	&	$-$0.08	&	0.39	&	0.35	&	0.47	&	0.10	&	0.07	&	$-$0.10	&	0.45	&	0.53	\\
2M17373336$-$0315103	&	E	&	$-$1.09	&	$-$1.09	&	$-$0.71	&	0.55	&	0.16	&	1.11	&	0.37	&	0.39	&	0.08	&	$-$0.21	&	...	&	...	\\
2M17373351$-$0314396	&	E	&	$-$1.16	&	$-$1.17	&	$-$0.49	&	0.61	&	0.23	&	1.39	&	0.45	&	0.35	&	0.02	&	$-$0.10	&	0.34	&	0.38	\\
2M17373366$-$0312495	&	P2	&	$-$1.11	&	$-$1.11	&	0.44	&	0.13	&	0.43	&	0.42	&	0.28	&	0.29	&	0.04	&	$-$0.09	&	0.40	&	0.41	\\
2M17373382$-$0315285	&	E	&	$-$1.17	&	$-$1.17	&	$-$0.18	&	0.59	&	0.14	&	1.04	&	0.40	&	0.37	&	0.11	&	$-$0.08	&	0.22	&	0.42	\\
2M17373390$-$0313291	&	P2	&	$-$1.23	&	$-$1.19	&	0.47	&	0.18	&	0.45	&	0.50	&	0.33	&	0.29	&	0.02	&	$-$0.17	&	0.23	&	0.31	\\
2M17373395$-$0313138	&	P2	&	$-$1.10	&	...	&	0.56	&	0.14	&	0.37	&	0.48	&	0.38	&	0.41	&	0.16	&	$-$0.11	&	0.39	&	0.35	\\
2M17373458$-$0315552	&	P2	&	$-$1.13	&	$-$1.13	&	0.63	&	0.04	&	0.45	&	0.41	&	0.22	&	0.33	&	0.11	&	$-$0.06	&	0.30	&	0.38	\\
2M17373460$-$0314057	&	P1	&	$-$1.17	&	$-$1.18	&	0.62	&	$-$0.01	&	0.34	&	0.58	&	0.38	&	0.35	&	0.05	&	0.00	&	0.30	&	0.31	\\
2M17373470$-$0315094	&	...	&	...	&	...	&	...	&	...	&	...	&	...	&	...	&	...	&	...	&	...	&	...	&	...	\\
2M17373606$-$0315251	&	P1	&	$-$1.21	&	$-$1.22	&	0.76	&	$-$0.11	&	0.43	&	0.42	&	0.21	&	0.30	&	0.14	&	$-$0.12	&	0.32	&	0.60	\\
2M17373691$-$0313304	&	E	&	$-$1.10	&	$-$1.11	&	$-$0.40	&	0.53	&	...	&	1.27	&	0.35	&	0.42	&	0.13	&	$-$0.06	&	0.33	&	0.36	\\
2M17373796$-$0314120	&	P1	&	$-$1.20	&	$-$1.21	&	0.90	&	$-$0.13	&	0.42	&	0.45	&	0.34	&	0.26	&	0.00	&	$-$0.11	&	0.28	&	0.39	\\
2M17373815$-$0314496	&	E	&	$-$1.15	&	$-$1.15	&	$-$0.15	&	0.58	&	0.27	&	1.15	&	0.36	&	0.37	&	0.21	&	$-$0.03	&	0.42	&	0.40	\\
2M17373864$-$0313445	&	P2	&	$-$1.25	&	$-$1.25	&	0.60	&	0.11	&	0.40	&	0.51	&	0.34	&	0.32	&	0.14	&	$-$0.04	&	0.25	&	0.40	\\
2M17373909$-$0313237	&	...	&	...	&	...	&	...	&	...	&	...	&	...	&	...	&	...	&	...	&	...	&	...	&	...	\\
2M17373921$-$0314264	&	P1	&	$-$1.20	&	$-$1.16	&	0.70	&	$-$0.10	&	0.37	&	0.15	&	0.29	&	0.29	&	$-$0.01	&	$-$0.17	&	0.20	&	0.27	\\
2M17373944$-$0314480	&	P1	&	$-$1.12	&	$-$1.14	&	...	&	0.00	&	0.44	&	0.37	&	0.17	&	0.25	&	$-$0.04	&	$-$0.11	&	0.30	&	0.28	\\
2M17373961$-$0315269	&	P1	&	$-$1.17	&	$-$1.17	&	0.82	&	$-$0.08	&	0.39	&	0.41	&	0.32	&	0.30	&	0.17	&	$-$0.05	&	0.32	&	0.52	\\
2M17374031$-$0315544	&	P2	&	$-$1.17	&	$-$1.17	&	0.62	&	0.18	&	0.39	&	0.54	&	0.31	&	0.34	&	0.24	&	$-$0.08	&	0.33	&	0.40	\\
2M17374053$-$0314592	&	P2	&	$-$1.17	&	$-$1.17	&	0.35	&	$-$0.04	&	0.34	&	0.47	&	0.30	&	0.21	&	0.03	&	$-$0.24	&	0.24	&	...	\\
2M17374093$-$0315327	&	P1	&	$-$1.01	&	$-$1.01	&	0.53	&	$-$0.18	&	0.30	&	0.29	&	0.17	&	0.31	&	0.05	&	$-$0.19	&	0.13	&	0.26	\\
2M17374103$-$0313167	&	P2	&	$-$1.13	&	$-$1.14	&	...	&	0.21	&	0.40	&	0.38	&	0.32	&	0.42	&	0.12	&	0.00	&	0.39	&	0.49	\\
2M17374139$-$0313528	&	...	&	...	&	...	&	...	&	...	&	...	&	...	&	...	&	...	&	...	&	...	&	...	&	...	\\
2M17374163$-$0314213	&	E	&	$-$1.05	&	$-$1.06	&	0.03	&	0.71	&	...	&	0.90	&	0.66	&	0.33	&	0.06	&	$-$0.10	&	0.18	&	0.31	\\
2M17374212$-$0315069	&	P1	&	$-$1.09	&	$-$1.13	&	0.82	&	$-$0.01	&	0.31	&	0.49	&	0.27	&	0.30	&	0.11	&	$-$0.15	&	0.40	&	0.46	\\
2M17374273$-$0315238	&	P2	&	$-$1.08	&	$-$1.10	&	0.43	&	0.20	&	0.30	&	0.58	&	0.35	&	0.35	&	0.13	&	$-$0.13	&	0.29	&	0.35	\\
2M17374309$-$0314044	&	P2	&	$-$1.15	&	$-$1.16	&	0.46	&	0.21	&	0.35	&	0.55	&	0.21	&	0.29	&	0.07	&	$-$0.16	&	0.31	&	0.41	\\
2M17374421$-$0313013	&	E	&	$-$1.01	&	$-$1.01	&	...	&	0.51	&	0.03	&	1.26	&	0.41	&	0.29	&	0.00	&	$-$0.23	&	0.35	&	0.31	\\
2M17374450$-$0314255	&	P1	&	$-$1.11	&	$-$1.12	&	0.66	&	$-$0.10	&	0.43	&	0.43	&	0.37	&	0.25	&	$-$0.02	&	$-$0.10	&	0.39	&	0.28	\\
2M17374518$-$0314040	&	E	&	$-$1.12	&	$-$1.13	&	$-$0.13	&	...	&	0.26	&	1.06	&	0.32	&	0.38	&	0.04	&	$-$0.14	&	0.25	&	0.28	\\
2M17374581$-$0314407	&	P1	&	$-$1.00	&	$-$1.03	&	0.63	&	0.00	&	0.32	&	0.43	&	0.28	&	0.28	&	0.02	&	$-$0.13	&	0.39	&	0.52	\\
2M17374610$-$0315079	&	P2	&	$-$1.13	&	$-$1.12	&	0.43	&	0.07	&	0.40	&	0.34	&	0.45	&	0.23	&	$-$0.04	&	$-$0.15	&	0.20	&	0.36	\\
  \hline
  \end{tabular}
\end{table}
\end{landscape}

\begin{landscape}
\begin{table}
  \caption{Table 3: abundance ratio uncertainties for all member stars.}
  \label{tab:errors}
  \begin{tabular}{ccccccccccccc}
  \hline
  Star	&	$\Delta$[Fe{\sevensize I}/H]	&	$\Delta$[Fe{\sevensize II}/H]	&	$\Delta$[O{\sevensize I}/Fe]	&	$\Delta$[Na{\sevensize I}/Fe]	&	$\Delta$[Mg{\sevensize I}/Fe]	&	$\Delta$[Al{\sevensize I}/Fe]	&	$\Delta$[Si{\sevensize I}/Fe]	&	$\Delta$[Ca{\sevensize I}/Fe]	&	$\Delta$[Cr{\sevensize I}/Fe]	&	$\Delta$[Ni{\sevensize I}/Fe]	&	$\Delta$[La{\sevensize II}/Fe]	&	$\Delta$[Eu{\sevensize II}/Fe]\\
  (2MASS)	&	(dex)	&	(dex)   &	(dex)   &	(dex)   &	(dex)   &	(dex)   &	(dex)   &	(dex)   &	(dex)   &	(dex)   &	(dex)   &	(dex)\\
  \hline
2M17372576$-$0314077	&	0.10	&	0.12	&	0.10	&	0.08	&	0.07	&	0.08	&	0.07	&	0.07	&	0.09	&	0.05	&	0.13	&	0.11	\\
2M17372832$-$0314210	&	...	&	...	&	...	&	...	&	...	&	...	&	...	&	...	&	...	&	...	&	...	&	...	\\
2M17373006$-$0314502	&	...	&	...	&	...	&	...	&	...	&	...	&	...	&	...	&	...	&	...	&	...	&	...	\\
2M17373048$-$0315530	&	...	&	...	&	...	&	...	&	...	&	...	&	...	&	...	&	...	&	...	&	...	&	...	\\
2M17373178$-$0313592	&	0.10	&	0.14	&	0.10	&	0.06	&	0.07	&	0.08	&	0.08	&	0.07	&	0.07	&	0.09	&	0.13	&	0.11	\\
2M17373225$-$0311152	&	0.10	&	0.12	&	0.10	&	0.11	&	0.07	&	0.07	&	0.10	&	0.06	&	0.08	&	0.05	&	0.13	&	0.11	\\
2M17373243$-$0313213	&	0.10	&	0.14	&	0.10	&	0.05	&	...	&	0.08	&	0.07	&	0.05	&	0.10	&	0.12	&	0.13	&	0.11	\\
2M17373305$-$0311386	&	0.10	&	0.13	&	...	&	0.17	&	0.07	&	0.07	&	0.12	&	0.08	&	0.09	&	0.07	&	0.13	&	0.11	\\
2M17373325$-$0315382	&	0.10	&	0.11	&	0.10	&	0.08	&	0.07	&	0.08	&	0.07	&	0.09	&	0.07	&	0.07	&	0.13	&	0.11	\\
2M17373336$-$0315103	&	0.10	&	0.12	&	0.10	&	0.05	&	0.07	&	0.16	&	0.10	&	0.06	&	0.10	&	0.08	&	...	&	...	\\
2M17373351$-$0314396	&	0.10	&	0.12	&	0.10	&	0.08	&	0.07	&	0.08	&	0.10	&	0.05	&	0.12	&	0.07	&	0.13	&	0.11	\\
2M17373366$-$0312495	&	0.10	&	0.13	&	0.10	&	0.06	&	0.07	&	0.09	&	0.08	&	0.06	&	0.07	&	0.07	&	0.13	&	0.11	\\
2M17373382$-$0315285	&	0.10	&	0.17	&	0.10	&	0.08	&	0.07	&	0.11	&	0.08	&	0.06	&	0.09	&	0.07	&	0.13	&	0.11	\\
2M17373390$-$0313291	&	0.10	&	0.12	&	0.10	&	0.07	&	0.07	&	0.07	&	0.07	&	0.06	&	0.07	&	0.06	&	0.13	&	0.11	\\
2M17373395$-$0313138	&	0.10	&	...	&	0.10	&	0.06	&	0.07	&	0.09	&	0.07	&	0.07	&	0.08	&	0.07	&	0.13	&	0.11	\\
2M17373458$-$0315552	&	0.10	&	0.13	&	0.10	&	0.07	&	0.07	&	0.08	&	0.07	&	0.06	&	0.07	&	0.07	&	0.13	&	0.11	\\
2M17373460$-$0314057	&	0.10	&	0.11	&	0.10	&	0.05	&	0.07	&	0.08	&	0.07	&	0.07	&	0.07	&	0.06	&	0.13	&	0.11	\\
2M17373470$-$0315094	&	...	&	...	&	...	&	...	&	...	&	...	&	...	&	...	&	...	&	...	&	...	&	...	\\
2M17373606$-$0315251	&	0.10	&	0.12	&	0.10	&	0.05	&	0.07	&	0.14	&	0.08	&	0.05	&	0.09	&	0.07	&	0.13	&	0.11	\\
2M17373691$-$0313304	&	0.10	&	...	&	0.10	&	0.05	&	...	&	0.08	&	0.09	&	0.07	&	0.08	&	0.08	&	0.13	&	0.11	\\
2M17373796$-$0314120	&	0.10	&	0.12	&	0.10	&	0.05	&	0.07	&	0.08	&	0.07	&	0.08	&	0.07	&	0.05	&	0.13	&	0.11	\\
2M17373815$-$0314496	&	0.10	&	0.12	&	0.10	&	0.07	&	0.07	&	0.07	&	0.07	&	0.07	&	0.07	&	0.06	&	0.13	&	0.11	\\
2M17373864$-$0313445	&	0.10	&	0.12	&	0.10	&	0.11	&	0.07	&	0.08	&	0.07	&	0.08	&	0.07	&	0.06	&	0.13	&	0.11	\\
2M17373909$-$0313237	&	...	&	...	&	...	&	...	&	...	&	...	&	...	&	...	&	...	&	...	&	...	&	...	\\
2M17373921$-$0314264	&	0.10	&	0.14	&	0.10	&	0.05	&	0.07	&	0.07	&	0.12	&	0.07	&	0.09	&	0.07	&	0.13	&	0.11	\\
2M17373944$-$0314480	&	0.10	&	0.11	&	...	&	0.09	&	0.07	&	0.08	&	0.07	&	0.08	&	0.09	&	0.06	&	0.13	&	0.11	\\
2M17373961$-$0315269	&	0.10	&	0.14	&	0.10	&	0.07	&	0.07	&	0.11	&	0.09	&	0.07	&	0.12	&	0.06	&	0.13	&	0.11	\\
2M17374031$-$0315544	&	0.10	&	0.12	&	0.10	&	0.05	&	0.07	&	0.10	&	0.07	&	0.08	&	0.07	&	0.06	&	0.13	&	0.11	\\
2M17374053$-$0314592	&	0.10	&	0.12	&	0.10	&	0.07	&	0.07	&	0.08	&	0.08	&	0.08	&	0.08	&	0.06	&	0.13	&	0.11	\\
2M17374093$-$0315327	&	0.10	&	0.11	&	0.10	&	0.06	&	0.07	&	0.07	&	0.09	&	0.06	&	0.08	&	0.07	&	0.13	&	0.11	\\
2M17374103$-$0313167	&	0.10	&	0.12	&	...	&	0.07	&	0.07	&	0.08	&	0.10	&	0.12	&	0.10	&	0.08	&	0.13	&	0.11	\\
2M17374139$-$0313528	&	...	&	...	&	...	&	...	&	...	&	...	&	...	&	...	&	...	&	...	&	...	&	...	\\
2M17374163$-$0314213	&	0.10	&	0.11	&	0.10	&	0.06	&	0.07	&	0.07	&	0.07	&	0.06	&	0.12	&	0.12	&	0.13	&	0.11	\\
2M17374212$-$0315069	&	0.10	&	0.12	&	0.10	&	0.06	&	0.07	&	0.07	&	0.09	&	0.07	&	0.11	&	0.05	&	0.13	&	0.11	\\
2M17374273$-$0315238	&	0.10	&	0.13	&	0.10	&	0.07	&	0.07	&	0.07	&	0.08	&	0.07	&	0.09	&	0.06	&	0.13	&	0.11	\\
2M17374309$-$0314044	&	0.10	&	0.13	&	0.10	&	0.05	&	0.07	&	0.13	&	0.07	&	0.05	&	0.13	&	0.06	&	0.13	&	0.11	\\
2M17374421$-$0313013	&	0.10	&	0.12	&	...	&	0.06	&	0.07	&	0.07	&	0.10	&	0.09	&	0.10	&	0.06	&	0.13	&	0.11	\\
2M17374450$-$0314255	&	0.10	&	0.12	&	0.10	&	0.06	&	0.07	&	0.08	&	0.07	&	0.08	&	0.08	&	0.08	&	0.13	&	0.11	\\
2M17374518$-$0314040	&	0.10	&	0.12	&	0.10	&	...	&	0.07	&	0.08	&	0.07	&	0.06	&	0.11	&	0.09	&	0.13	&	0.11	\\
2M17374581$-$0314407	&	0.10	&	0.11	&	0.10	&	0.06	&	0.07	&	0.07	&	0.10	&	0.06	&	0.08	&	0.06	&	0.13	&	0.11	\\
2M17374610$-$0315079	&	0.10	&	0.12	&	0.10	&	0.08	&	0.07	&	0.07	&	0.10	&	0.08	&	0.09	&	0.08	&	0.13	&	0.11	\\
  \hline
  \end{tabular}
\end{table}
\end{landscape}

\begin{landscape}
\begin{table}
  \caption{Table 4: mean abundance ratios and Welch's \emph{t}-test results for
the P1, P2, and E populations.}
  \label{tab:populations}
  \begin{tabular}{cccccccccccc}
  \hline
  \multicolumn{12}{c}{Mean Abundance Ratios}\\
  \hline
Population	&	[Fe/H]	&	[O/Fe]	&	[Na/Fe]	&	[Mg/Fe]	&	[Al/Fe]	&	[Si/Fe]	&	[Ca/Fe]	&	[Cr/Fe]	&	[Ni/Fe]	&	[La/Fe]	&	[Eu/Fe]\\
...	&	(dex)	&       (dex)	&       (dex)	&       (dex)	&       (dex)	&       (dex)	&       (dex)	&       (dex)	&       (dex)	&       (dex)	&       (dex)\\
  \hline
P1	&	$-$1.13	&	0.69	&	$-$0.08	&	0.38	&	0.39	&	0.30	&	0.27	&	0.04	&	$-$0.11	&	0.31	&	0.39	\\
P2	&	$-$1.15	&	0.49	&	0.14	&	0.39	&	0.49	&	0.32	&	0.32	&	0.09	&	$-$0.11	&	0.29	&	0.37	\\
E	&	$-$1.11	&	$-$0.28	&	0.58	&	0.20	&	1.16	&	0.44	&	0.36	&	0.08	&	$-$0.12	&	0.28	&	0.36	\\
  \hline
  \multicolumn{12}{c}{\emph{t}-statistics, degrees of freedom, and p-values}\\
  \hline
\emph{t} (P1-P2)	&	0.50	&	4.09	&	$-$7.40	&	$-$0.83	&	$-$2.50	&	$-$0.64	&	$-$2.11	&	$-$1.76	&	0.23	&	0.67	&	0.47	\\
d.o.f	&	18	&	20	&	23	&	22	&	20	&	19	&	24	&	24	&	24	&	23	&	16	\\
p-value	&	0.62	&	5.50$\times$10$^{\rm -4}$	&	1.68$\times$10$^{\rm -7}$	&	0.42	&	0.02	&	0.53	&	0.05	&	0.09	&	0.82	&	0.51	&	0.65	\\
\hline
\emph{t} (P1-E)	&	$-$0.89	&	10.62	&	$-$23.66	&	4.65	&	$-$13.03	&	$-$2.91	&	$-$3.81	&	$-$1.16	&	0.52	&	0.77	&	0.65	\\
d.o.f	&	19	&	10	&	15	&	8	&	14	&	14	&	19	&	18	&	15	&	15	&	16	\\
p-value	&	0.38	&	1.17$\times$10$^{\rm -6}$	&	3.54$\times$10$^{\rm -13}$	&	1.65$\times$10$^{\rm -3}$	&	3.92$\times$10$^{\rm -9}$	&	0.01	&	1.19$\times$10$^{\rm -3}$	&	0.26	&	0.62	&	0.45	&	0.53	\\
  \hline
\emph{t} (P2-E)	&	$-$1.64	&	8.58	&	$-$13.69	&	5.16	&	$-$12.43	&	$-$2.74	&	$-$1.45	&	0.53	&	0.31	&	0.24	&	0.29	\\
d.o.f	&	15	&	9	&	19	&	7	&	11	&	11	&	21	&	19	&	17	&	13	&	17	\\
p-value	&	0.12	&	1.14$\times$10$^{\rm -5}$	&	2.68$\times$10$^{\rm -11}$	&	1.10$\times$10$^{\rm -3}$	&	1.01$\times$10$^{\rm -7}$	&	0.02	&	0.16	&	0.61	&	0.76	&	0.81	&	0.78	\\
  \hline
  \end{tabular}
\end{table}
\end{landscape}


%
%


\bsp	
\label{lastpage}
\end{document}